\documentclass{article} 
\usepackage{iclr2025_conference,times}

\usepackage{amsmath,amsfonts,bm}

\def\eqref#1{equation~\ref{#1}}

\def\1{\bm{1}}

\DeclareMathAlphabet{\mathsfit}{\encodingdefault}{\sfdefault}{m}{sl}
\SetMathAlphabet{\mathsfit}{bold}{\encodingdefault}{\sfdefault}{bx}{n}

\usepackage{hyperref}
\usepackage{url}

\usepackage{adjustbox}
\usepackage{algorithmic}
\usepackage{algorithm}
\usepackage{booktabs}
\usepackage{cleveref}
\usepackage{enumitem}
\usepackage{pifont}
\usepackage{placeins}
\usepackage{siunitx}
\usepackage{soul}
\usepackage{subcaption}
\usepackage{tikz}
\usepackage{wrapfig}
\usepackage{xcolor}         
\usepackage{xspace}

\newcommand{\codexglue}{\textsc{codexglue}}
\newcommand{\codetp}{\text{CodeT5+ \xspace}}
\newcommand{\codesearchnet}{\text{CodeSearchNet}}
\newcommand{\pyonefifty}{\text{PY150}}
\newcommand{\javacorpus}{\text{JavaCorpus}}

\definecolor{green}{RGB}{66, 180, 20}
\definecolor{red}{RGB}{219,68,55}
\definecolor{blue}{RGB}{66,133,244}
\definecolor{magenta}{RGB}{200, 90, 200}

\usepackage{float}

\newfloat{algo}{t}{lop}
\floatname{algo}{Algorithm}

\let\oldnl\nl
\newcommand{\nonl}{\renewcommand{\nl}{\let\nl\oldnl}}

\newcommand{\tool}{{\sc NoEsis}\xspace}
\newcommand{\cmark}{\ding{51}}%
\newcommand{\xmark}{\ding{55}}%

\newif\ifcomment
\commenttrue

\ifcomment
\newcommand{\steve}[1]{\sethlcolor{cyan}\hl{[\textbf{Steve:} #1]}}
\newcommand{\rob}[1]{\sethlcolor{orange}\hl{[\textbf{Rob:} #1]}}
\else
\newcommand{\steve}[1]{}
\newcommand{\rob}[1]{}
\fi

\newcommand{\blackcircle}[1]{%
    \tikz[baseline=(char.base)]{
        \node[shape=circle, draw=black, fill=black, text=white, inner sep=0.5mm] (char) {\textbf{#1}};
    }%
}

\title{NoEsis: Differentially Private Knowledge Transfer in Modular LLM Adaptation}

\author{Rob Romijnders\thanks{Work done during internship at Brave Research.}, Stefanos Laskaridis, Ali Shahin Shamsabadi \& Hamed Haddadi  \\
    Brave Research
}

\iclrfinalcopy 
\begin{document}

\maketitle

\begin{abstract}
Large Language Models (LLM) are typically trained on vast amounts of data from various sources. Even when designed modularly (e.g., Mixture-of-Experts), LLMs can leak privacy on their sources.
Conversely, training such models in isolation arguably prohibits generalization.
To this end, we propose a framework, \tool, which builds upon the desired properties of \textit{modularity}, \textit{privacy}, and \textit{knowledge transfer}. \tool\ integrates differential privacy with a hybrid two-staged parameter-efficient fine-tuning that combines domain-specific low-rank adapters, acting as experts, with common prompt tokens, acting as a knowledge-sharing backbone.
Results from our evaluation on \codexglue\ showcase that \tool\ can achieve provable privacy guarantees with tangible knowledge transfer across domains, and empirically show protection against Membership Inference Attacks. Finally, on code completion tasks, \tool\ bridges at least 77\% of the accuracy gap between the non-shared and the non-private baseline.
\end{abstract}
\section{Introduction} \label{sec:introduction}

Large Language Models have brought much disruption in the field of Artificial Intelligence and have transformed various use-cases, from intelligent assistants~\citep{dong2023towards} and code co-pilots~\citep{human_eval} to agentic web browsing~\citep{zhenggpt} and enhanced tutoring~\citep{kotalwar2024hints}. They have shown great scaling potential, devouring terabytes of raw textual or multi-modal data~\citep{kaplan2020scaling} without their performance plateauing. As this trend continues, all public resources will eventually be consumed. 
Therefore, tapping into private data silos will become the next significant source of information~\citep{shumailov2024ai,iacob2024worldwide}.

However, copyright~\citep{xu-etal-2024-llms} and privacy laws~\citep{eurlex2024,ccpa} may prevent models from being trained and served as-is, uniformly across the globe. 
This introduces the need to orchestrate model training that is somehow separated per region or source. Maintaining separate models, though, quickly becomes intractable and burdensome.
Private organizations can own data they want to use for their custom LLM but not expose it publicly~\cite {carlini_ref_model_MIA, openai2023gpt35}. For instance, client institutions may wish to train domain-specific Copilots~\citep{githubcopilot} without leaking proprietary information~\citep{291327} to the public domain.

\looseness=-1 To approach this problem, we draw from Modular Learning~\citep{pfeiffer2023modular} for routing knowledge across parts of a neural network and adaptively serve to different domains. While off-the-shelf 
Mixture-of-Experts (MoE)~models~\citep{Cai2024ASO} adopt an architecture where different domains can share common parameters --
thus enabling knowledge transfer. However, they can introduce privacy risks~\citep{carlini2019secret} exactly because of this sharing.
In addition, training an entire MoE model under Differential Privacy (DP) significantly reduces its utility as training a large shared backbone network over multiple domains requires adding large amounts of DP noise. 
Therefore, there is an inherent tension between \textit{modularity}, \textit{privacy} and \textit{knowledge sharing}.

\looseness=-1 We propose a privacy-friendly architecture for differentially private modular learning, Nexus-of-Experts (NoE). This inherits a domain-routed mixture of low-rank adapters model (Mix-LoRA), applied on the fully connected layer of each transformer block, further enhanced by shared tokens obtained through a DP training algorithm so that it can be activated across domains (Fig.~\ref{fig:noesis_pipeline}). This architecture allows for a modular architecture and tunably shares knowledge between domains privately. 
Compared to conventional MoE methods, we enable efficient \mbox{document-private} learning. Compared to adaptive parameter-efficient fine-tuning (PEFT) alternatives, we facilitate knowledge transfer between domains while providing modularity at deployment time.

In summary, our contributions are the following:
\begin{itemize}[noitemsep,topsep=0pt,leftmargin=10mm]
\item We expose privacy leakage on existing routing-based models, showcasing the need for privacy-preserving training in these models.
\item We propose \tool, a domain-routed, hybrid PEFT solution that adopts a shared DP-trained set of prompt tokens that act as a knowledge-sharing backbone and enable knowledge transfer between domains, and a \mbox{Mix-LoRA} that act as private domain experts.
\item \looseness=-1 We analyze the balance between privacy and model expressiveness by tuning the number of trainable parameters and the noise needed to ensure privacy. Indicatively, our hybrid method performs up to \mbox{1.4\%} accuracy points better than the non-hybrid, LoRA-only variant.
\item We apply our technique to the problem of code generation, adopting different programming languages as our domains and ensuring privacy at a document level. Our results showcase that \tool\ can achieve from 1.4 to 4.9\% points better downstream accuracy, averaged across domains, while protecting our model against privacy leakage.
\end{itemize}
\vspace{-3mm}
\section{Desiderata}\label{sec:desiderata}
\vspace{-2mm}
We study the problem of fine-tuning an LLM for various private domains.
This section explains three properties of the problem and shows how \tool\ uniquely satisfies all of them (Table~\ref{tab:venn_diagram}).

\begin{table}[t]
\centering
\caption{Learning approaches and deployment characteristics of each solution. \tool\ is the only method that is \textit{modular}, \textit{private} and can benefit from \textit{knowledge transfer} among domains.}
\label{tab:venn_diagram}
\vspace{-0.2cm}
\setlength{\tabcolsep}{4pt}
\begin{adjustbox}{max width=\textwidth}
\begin{tabular}{c c c | l l}
\toprule
 Modular & Private & Transfer & Approach & Remarks  \\
\midrule
\xmark & \xmark & \cmark & Monolithic model~\citep{codet5p} & Leaking privacy in model parameters \\
\xmark & \cmark & \cmark & Private learning~(\citep{codet5p}+\citep{abadi_dpsgd}) & Low performance \\
\cmark & \cmark & \xmark & MoE (share-nothing)~\citep{zhou2024moe} & No benefit from knowledge transfer \\
\cmark & \xmark & \cmark & MoE (shared-expert)~\citep{multicoder} & Privacy leakage in shared expert \\
\midrule
\cmark & \cmark & \cmark & \tool\ (ours) & Addresses all three problems \\
\bottomrule
\end{tabular}
\end{adjustbox}
\vspace{-0.3cm}
\end{table}

\noindent
\textbf{Knowledge Transfer.} 
When training on multiple data sources, knowledge transfer is the improvement in results when learning a joint model from all sources, instead of learning a separate model per source~\citep{ruder-etal-2019-transfer}. 

In our multi-domain setting, we desire more accurate predictions in each domain while maintaining privacy regarding potential adversaries from other domains.

\noindent
\textbf{Modular Learning.} Modular learning refers to models whose execution subgraphs can be specialized in terms of function~\citep{pfeiffer2023modular,Cai2024ASO,switch_transformer}.
The benefits of modularity for our problem are two-fold. 
Computationally, modularity allows us to train once and serve model components everywhere without maintaining multiple copies or replicating knowledge.
Privacy-related, separating domain-specific and shared parameters allows for a clear separation of concerns in terms of domains. This can be important for serving under different jurisdictions and providing formal guarantees about the flow of information.

\noindent
\textbf{Preserving privacy.}
Machine learning models can inadvertently leak sensitive training data~\citep{carlini_ref_model_MIA}, which raises privacy concerns when private information is involved. 
We aim to improve model accuracy by training on multiple private data sets from different domains without impacting their privacy.
Therefore, we formalize a threat model when training a model on private datasets from multiple domains, where an adversary in one domain should not be able to discern information about training data in another domain.

These requirements are essential in various real-world scenarios. An example is a globally operating organization that wants to train a single model but serve modularly across different regions and ensure that the model does not leak information between regions due to various copyright or privacy laws~\citep{apple_eu, meta_eu}. 
Finally, the modular aspect can be helpful in scenarios where an organization wants to train a model on multiple domains, such as legal, instructional, and administrative documents~\citep{salesforce2024agentforce}, but wants to ensure that the model does not leak information between domains.\looseness=-1

\vspace{-1mm}
\section{\tool}\label{sec:method}
\vspace{-2mm}
\looseness=-1 Aiming for the properties of \textit{modularity}, \textit{knowledge transfer}, and \textit{privacy}, we introduce \tool, a hybrid, DP-trained, domain-routed Nexus-of-Experts (NoE) solution that adopts a shared set of trainable tokens and Mix-LoRA experts and enables knowledge transfer between domains. 
%
%

\textbf{Domain-routed models.} \tool\ is fundamentally a modular architecture that enables the specialization of network components to different domains. Given dataset $D_i^{(k)}$ from domain $k$, where $k \in [1, K]$, we route the data from each domain to the respective expert. Instead of operating directly on the backbone model parameters, we use a Mix-LoRA~\citep{hulora,mixlora_li_2024} architecture that keeps the backbone network frozen and only tunes the adapters per domain. The adapters are applied on top of feed-forward network (FFN) layers of the transformer blocks to enable the ``logic'' of the network to be tuned per domain~\cite{mixlora_li_2024}. Mix-LoRA has been shown to be effective in routing-based neural networks~\citep{mixlora_li_2024,mixlora_domain,mixlora_iclr,mixlora_multimodal}.
Formally, let $W \in \mathbb{R}^{p \times q}$ be the weight matrix of a linear layer
in the model with parameters $\theta$. We decompose
\begin{equation}\label{eqn:mix-lora}
    W = W + \alpha \sum_{k=1}^K \underline{B^{(k)}}\underline{A^{(k)}},
\end{equation}
where $A^{(k)} \in \mathbb{R}^{r \times q}$ and $B^{(k)} \in \mathbb{R}^{p \times r}$ are the trainable (signified by the underline) low-rank matrices for the \( k \)-th domain-specific adapter, out of $K$ domains. 
The experimental section refers to the rank $r$ as the \textit{adapter rank}. $\alpha \in \mathbb{R}$ is a scalar constant to reduce the variance impact of the adapter. 

\begin{figure*}[t]
\centering
\vspace{-0.2cm}
\includegraphics[width=0.99\linewidth]{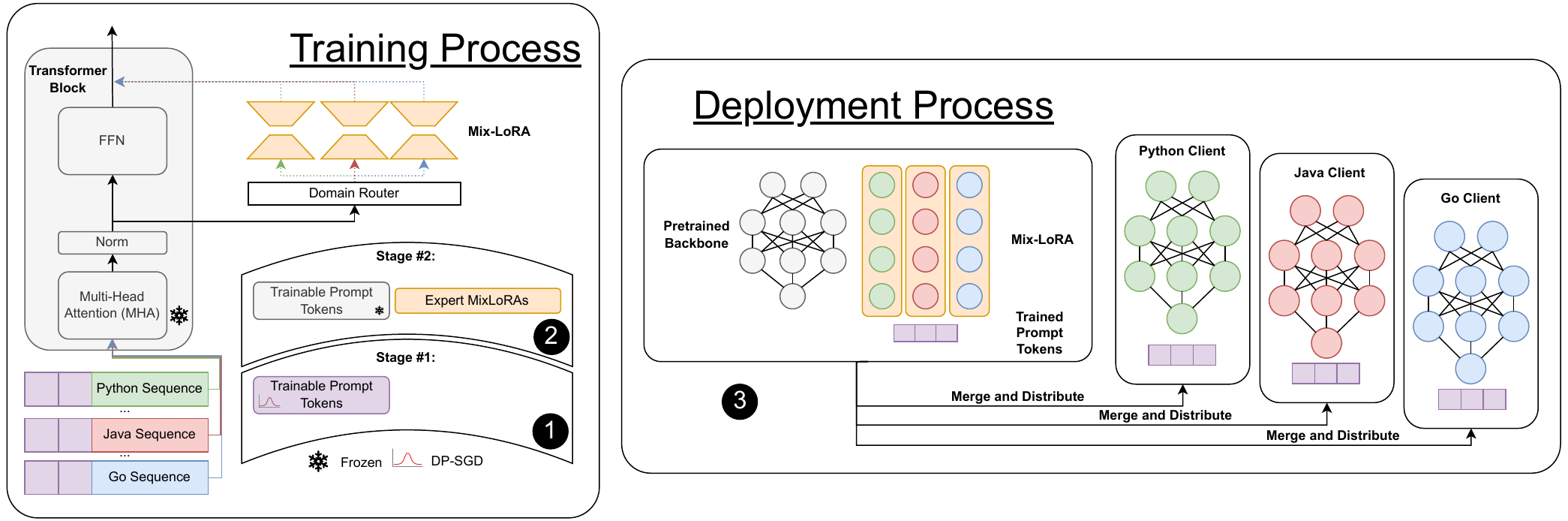}
\vspace{-1mm}
\caption{The training and the deployment process of \tool. The training process consists of two stages: Stage 1: The training of private prompt token parameters across domains; Stage 2: The training of expert Mix-LoRA per domain. The deployment involves merging the LoRA parameters with the backbone and sharing the privately trained prompt tokens with each downstream client.}
\label{fig:noesis_pipeline}
\vspace{-0.2cm}
\end{figure*}

\textbf{Enabling Knowledge Transfer.} 
We adopt a differentially private prompt-tuning mechanism to facilitate knowledge transfer between domains while keeping the number of shared trainable parameters minimal. Prompt tuning has emerged as a promising method for efficiently fine-tuning LLMs by introducing differentiable prompts, which are real-valued tokens that can be learned through back-propagation~\citep{li2021prefix,lester2021power}. Prior work demonstrates that DP prompt-tuning can achieve substantial predictive accuracy even with stringent privacy budgets, such as $\varepsilon < 1$~\citep{dp_prompt_pate}. By training only the differentiable prompts, we reduce the number of trainable parameters, thereby minimizing the required noise for DP.

\looseness=-1  Specifically, given input $X=\{x_1^{(k)}, x_2^{(k)},\dots, x_{L}^{(k)}\}$ of $L$ tokens, from domain $k\in K$, a common set of trainable prompt embeddings $P \in \mathbb{R}^{n_{pt} \times d_\text{vocab}}$ is prepended to the input sequence, $X^{\prime} = [P; X]$. $n_{pt}$ is the number of prompt tokens and $d_\text{vocab}$ the output dimensionality of the vocabulary. Differentially Private Stochastic Gradient Descent (DP-SGD)~\citep{dp_prompt_pate} tunes prompts under the framework of DP by clipping the per-sample gradient to a fixed norm and adding calibrated Gaussian noise to these clipped gradients before applying to update prompts. Concretely, this means $P^{(t+1)} \leftarrow P^{(t)} - \eta \tilde{g}_P^{(t)}$, where $\tilde{g}_P^{(t)}$ is the noisy clipped gradient with respect to the prompt tokens at step $t$ and $\eta$ is the step size.


\noindent
The \textbf{training Process} of \tool is illustrated in~\Cref{fig:noesis_pipeline} (left) and outlined in Algorithm~\ref{alg:mixed_priv_dpsgd} in the Appendix. It follows a two-stage procedure of private learning for the \textit{shared parameters} (Step \blackcircle{1}) and non-private learning for the \textit{domain-specific adapters} (Step \blackcircle{2}). In each iteration of the first stage, a batch of documents is sampled, and the gradients are computed for the domain-specific and shared parameters. The gradients are clipped and noised according to DP-SGD~\citep{abadi_dpsgd}. In the second stage, SGD refers to Stochastic Gradient Descent. During deployment (Step \blackcircle{3}), the LoRA parameters are sent to each respective domain and merged with the backbone.

\noindent
For \textbf{document-level DP,} we use the definition from~\citet{dwork_dp}. A randomized algorithm $f(\cdot)$ is ($\varepsilon,\delta$)-differentially private if the following holds for any two adjacent data sets $D$, $D'$, and for any subset $\mathcal{S}$ of outputs: $\Pr[f(D) \in \mathcal{S}] \leq e^\varepsilon \Pr[f(D') \in \mathcal{S}] + \delta$. Here, $D = \{ d_i \}_{i=1}^N$ is a dataset of $N$ documents. Each $d_i$ is a string of tokens of arbitrary length. In the multi-domain setting, a dataset comprises all documents in all domains. Two datasets $D$ and $D'$ are adjacent when they are identical except for one document in any domain being removed. For this privacy protection, we can calculate the noise multiplier as a function of total dataset size $N$ and batch size $N_b$ in Algorithm~\ref{alg:mixed_priv_dpsgd}.
\looseness=-1 In keeping with prior work, we run all privacy-focused experiments with $\varepsilon=1.0$ and $\delta = 10^{-6}$,
which is smaller than the rule-of-thumb of one divided by dataset size~\citep{choosing_epsilon}. 

\vspace{-1mm}
\section{Evaluation}
\vspace{-2mm}
Our evaluation adopts the task of code completion among different programming languages, which we consider private domains. First, we evaluate the knowledge transfer between domains. Second, we perform a sensitivity analysis for the number of trainable parameters. Finally, we run a privacy attack to investigate privacy leakage empirically.

\noindent
\textbf{Model.} We use a decoder-only model architecture for our experiments~\citep{radford2018improving} -- specifically, the pre-trained \codetp (220M) model. Given the previous tokens as input, the model is trained to autoregressively predict the next token in a sequence by using the softmax as likelihood and cross-entropy as the loss function. This setup is particularly suitable for programming language tasks, where the goal is to generate code one token at a time~\citep{multicoder}.

\noindent
\textbf{Datasets.} We use the Python, Java, and Go programming languages for our multi-domain training because the pre-trained model was trained on data specifically deduplicated for datasets in these languages~\citep{codet5p}. The deduplication is necessary to simulate private domains for fine-tuning. The dataset for Python is the \pyonefifty\ dataset~\citep{py150} and for Java the \javacorpus~\citep{javagc}, which follows a similar setting as~\citet{multicoder}. The Go data comes from the~\codesearchnet\ dataset~\citep{codesearchnet}. More details about the datasets can be found in Appendix~\ref{app:dataset}.

\begin{table}[t]
\centering
\setlength{\tabcolsep}{4pt}
\vspace{-0.2cm}
\caption{The main experimental comparison on modularity, privacy, and knowledge transfer. \tool\ uniquely addresses all three aspects, achieving high accuracy while obtaining DP at $\varepsilon = 1.0$ and enabling knowledge transfer across domains. (Knowledge transfer is defined as the increase in accuracy compared to ``Share Nothing,'' which does not have shared parameters between domains.)}
\vspace{-0.2cm}
\adjustbox{max width=\textwidth} {
\begin{tabular}{l l | c | c c c | c | c c c}
\toprule
& \textbf{Model} & \textbf{$\varepsilon$} & \textbf{Modular} & \textbf{Private} & \textbf{Transfer} & \textbf{PEFT} & \textbf{Python} & \textbf{Java} & \textbf{Go} \\
\midrule
\textit{(i)} & Share Nothing & 0.0 & \cmark & \cmark & \xmark & \cmark &  68.31 & 60.19 & 64.17 \\
\textit{(ii)} & Solo (separate models) & 1.0 & \xmark & \cmark & \xmark & \xmark & 30.19 & 14.84 & 4.84 \\
\textit{(iii)} & Monolithic Fine-tuning~\citep{abadi_dpsgd} & 1.0 & \xmark & \cmark & \cmark & \xmark & 36.05 & 23.54 & 18.34 \\  
\textit{(iv)} & Common LoRA Adapter (rc=512)$^*$~\citep{yu2021large} & 1.0 & \xmark & \cmark & \cmark & \cmark & 48.21 & 36.16 & 27.24  \\ %
\textit{(v)} & Prompt-Tuning Only (pt32)$^\dagger$~\citep{dp_prompt_pate} & 1.0 & \xmark & \cmark & \cmark & \cmark & 35.03 & 24.29 & 9.67  \\ %
\textit{(vi)} & \tool\ (pt32)$^\dagger$ & 1.0 & \cmark & \cmark & \cmark & \cmark & \textbf{69.14} & \textbf{61.18} & \textbf{66.53}  \\
\bottomrule
\multicolumn{8}{l}{{\small $^*$rc, rank $r$ of the common LoRA; $^\dagger$\textit{pt: number of trainable prompt tokens}}}
\end{tabular}
}
\vspace{-0.5cm}
\label{tab:mainresult}
\end{table}

\noindent
\textbf{Training Setup \& Hyperparameters.} 
We train our models on 4$\times$4090 GPUs, using a learning rate of $10^{-3}$. We use the privacy-accountant of the Opacus library to calculate the noise-multiplier constant~\citep{opacus}. An epoch is defined as sampling one 512-token block for each document in the dataset. This ensures optimal use of compute resources and maintains the privacy guarantee as each epoch uses a document exactly once. Details about our hyperparameter values can be found in Appendix~\ref{app:hyperparameters}.
{The code to run the experiments corresponding to Table~\ref{tab:mainresult} can be found at :  \href{https://github.com/RobRomijnders/noesis/}{github.com/RobRomijnders/noesis/}.}

\noindent
We compare \tool\ against various trainable baselines, namely:
\textit{i)}~\textit{Share Nothing}: a modular model without shared parameters, which is to simulate the effect of not having knowledge transfer;
\textit{ii)}~\textit{Solo}: a separate model fine-tuned separately for each domain;
\textit{iii)}~\textit{Monolithic Fine-tuning}: DP fine-tuning the whole model without any routable components~\citep{abadi_dpsgd}, which is to compare with Solo and confirm the effect of knowledge transfer in monolithic models~\citep{wmt_knowledge_transfer};
\textit{iv)}~\textit{Single Common Adapter}: A common LoRA trained with DP across domains~\citep{yu2021large}, which is to compare the influence of using PEFT instead of monolithic fine-tuning;
\textit{v)}~\textit{Prompt-Tuning Only}: tuning only the prompts under the DP guarantee, which is to investigate the impact of knowledge transfer without any modularity in the form of Mix-LoRA~\citep{dp_prompt_pate}.
Details about the baselines can be found in Appendix~\ref{app:baselines}. \looseness=-1

\begin{figure}[t]
\centering
\begin{minipage}[t]{0.59\textwidth}
\centering \vspace{-0.01\baselineskip}
\setlength{\tabcolsep}{4pt}
\captionof{table}{Prompt-tuning achieves the best trade-off between the number of shared parameters and accuracy, across domains. All results obtained under the privacy guarantee.}
\vspace{-0.2cm}
\begin{tabular}{l | c | l l l}
\toprule
\textbf{Model}  & \textbf{\# Shared Params.}  & \textbf{Python} & \textbf{Java} & \textbf{Go} \\
\midrule
rc64  & $7680 \times 768=5.9$M & 68.73 &  60.32 & 65.32 \\  
rc4  & $\ 480 \times 768=369$k & 68.77 & 60.45 & 65.39 \\ 
rc1 & $\ 120 \times 768=92$k & 68.74 & 60.43 & 65.26 \\ 
\midrule
pt120  & $120 \times 768=92$k & 69.13 & 61.09 & 66.49 \\
pt32  & $\ 32 \times 768=25$k & \textbf{69.14} & \textbf{61.18} & \textbf{66.53}  \\
pt8 &  $\ \ 8 \times 768=6$k & 69.13 & 61.16 & 66.49 \\
\bottomrule
\multicolumn{5}{l}{{\small rc, rank $r$ of the common LoRA; pt, number of prompt tokens}}
\end{tabular}
\vspace{-5mm}
\label{tab:common_params}
\end{minipage}\hfill
\begin{minipage}[t]{0.39\textwidth}
\centering \vspace{-0.01\baselineskip}
\includegraphics[width=\linewidth]{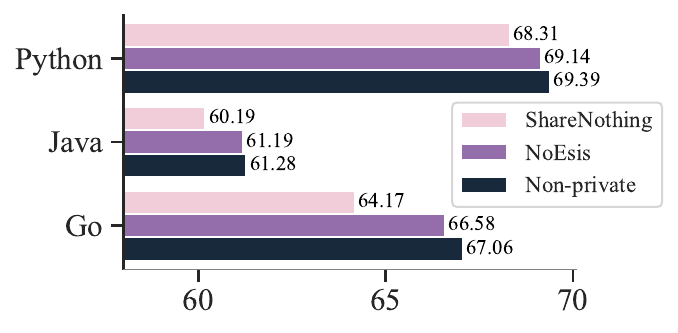}
\vspace{-5mm}
\captionof{figure}{Between the results of a non-shared model, which is the baseline, and a non-private model, which obtains the highest accuracy, \tool\ bridges the accuracy gap by more than 77\%.}
\label{fig:bridge}
\vspace{-5mm}
\end{minipage}
\end{figure}
\textbf{Experimental Results.} The results of comparing \tool\ against baselines are shown in Table~\ref{tab:mainresult}. Compared to Share Nothing, we see that \tool\ achieves an improvement of 0.83\% points on Python, 1.0\% points on Java, and 2.41\% points on Go. This indicates that knowledge transfer is beneficial for next token prediction, especially in the case of a scarce domain, such as Go, which has 50$\times$ less data than Python. Appendix~\ref{app:examples} shows qualitative examples of this comparison. To aid interpretation of numerical results, Appendix Figure~\ref{fig:dp_random_seeds} shows the results of \tool\ run on five different random seeds, making clear that generally the standard deviation is around 0.02\% point.

Comparing \tool\ to Prompt-Tuning Only, we observe an improvement of more than 30\% points across the three domains. This shows the benefit of using modularity as a privacy-friendly architecture. In general, the modular models achieve higher accuracy than the non-modular models, indicating the importance of modularity in multi-domain training. Comparing the experiments of Solo and Monolithic, we observe the effect of positive knowledge transfer in the non-modular setting, as the latter model achieves higher accuracy than each of the solo models. Finally, using a Common LoRA Adapter achieves about nine accuracy points improvement compared to monolithic fine-tuning, which corroborates earlier studies on PEFT and DP~\citep{yu2021dpLLM,yu2021large}.

Investigating the effect of knowledge transfer, we ask \textit{``how many shared parameters are optimal for knowledge transfer?''} This is contended by three directions: 
having too many parameters is vulnerable to \textit{overfitting}; having too few can lead to \textit{underfitting}; finally, more shared parameters require more noise in algorithm that satisfy DP~\citep{abadi_dpsgd}, which negatively influences learning. To investigate this contention, we train with $4\times$ more and $4\times$ less trainable prompt tokens. We also evaluate a variant with a LoRA as the trainable parameters for knowledge-sharing (explained in Appendix~\ref{app:noesis_rc}). 
From Table~\ref{tab:common_params}, the model with 32 shared prompt tokens (pt32) achieves the highest accuracy across all domains. This indicates that a moderate number of shared parameters appears optimal for knowledge transfer. Using a LoRA as a common parameter generally has lower accuracy than any prompt-tuning method. \emph{This indicates that prompt-tuning is a parameter-efficient option for providing knowledge transfer under differential privacy.}

Finally, we quantify the ``cost of privacy'' by comparing \tool with the same setup without the DP guarantee. As shown in Fig.~\ref{fig:bridge}, between the Share Nothing baseline (non-shared) and the non-private variant, \tool\ is able to achieve 84\% of the way on average. \emph{This indicates that \tool\ effectively achieves knowledge transfer and high accuracy while guaranteeing privacy.}

\textbf{Empirical Assessment of Privacy.} Complementing the analytical privacy guarantees, we measure empirical privacy based on a Membership Inference Attack (MIA)~\citep{shokri2017membership}. The attack aims to determine whether an input sequence, ${s}$, is a member of the model's training data. Several MIAs on LLMs have been proposed~\citep{camia,carlini_ref_model_MIA}, and we leverage an MIA based on the predictive log-likelihood scores and extend the attack to a multi-domain setting, named~\textit{cross-domain MIA}. The attack is particularly relevant for modular models~\citep{gshard,multicoder,gururangan-etal-2022-demix,zhou2024moe,komatsuzaki2022sparse,mixlora_iclr}, where shared parameters can leak information between domains.

\begin{wrapfigure}{hr}{6.9cm}
\centering
\vspace{-1mm}
\setlength{\tabcolsep}{3pt}
\captionof{table}{Mix-LoRA models have a privacy vulnerability in parameters that are shared between domains. \tool\ reduces this vulnerability while maintaining good predictive accuracy (Figure~\ref{fig:bridge}). The results are formatted as $\text{`non-private result'} \rightarrow \text{`private result'}$.}
\vspace{-0.2cm}
\begin{tabular}{l | r  l || r  l}
\toprule
\textbf{Attack} & \multicolumn{2}{c||}{AUC (\%)} & \multicolumn{2}{c}{TPR@1 (\%)}  \\
\midrule
Python (via Java) & 51.9 $\rightarrow$ & 50.6 & 1.1 $\rightarrow$ & 1.0  \\
Python (via Go) & 51.7 $\rightarrow$ & 50.7 & 1.2 $\rightarrow$ & 1.1  \\
\midrule
Java (via Go) & 65.4 $\rightarrow$ & 55.7 & 2.6 $\rightarrow$ & 1.4  \\ 
Java (via Python) & 64.8 $\rightarrow$ & 55.0 & 2.3 $\rightarrow$ & 1.3  \\ 
\midrule
Go (via Python) & 53.2 $\rightarrow$ & 50.6 &  1.1 $\rightarrow$ & 0.7 \\
Go (via Java) & 54.0 $\rightarrow$ & 50.1 & 1.3 $\rightarrow$ & 0.6 \\
\bottomrule
\end{tabular}
\vspace{-3mm}
\label{tab:roc_mia}
\end{wrapfigure}

The cross-domain attack follows a threat model where an attacker can make predictions using the parameters in a particular domain and aims to infer the membership of training data in another domain.
The scoring function $S(\mathbf{s}) = \mathcal{L}(\mathbf{s};\mathcal{M})$ is the average log-likelihood assigned to each successive token. The attacker compares this to a threshold $\tau$ to predict the membership.
The attack success is measured with Area Under the ROC Curve (AUC) and the True Positive Rate (TPR) -- defined in Appendix~\ref{app:mia-details}. The results in Table~\ref{tab:roc_mia} demonstrate that the original Mix-LoRA models can leak private information. For example, attacking Java via the Go domain, \tool\ has less empirical privacy leakage, measured by AUC, from 65.4 to 55.7\% for \tool. Following~\citet{MIA_argues_tpr}, Table~\ref{tab:roc_mia} reports the TPR at a 1\% False Positive Rate (FPR), which indicates attack success when the attacker makes only 1\% false positives. Across all domains, the attack on \tool\ has a lower (better) TPR compared to a Mix-LoRA model.

\vspace{-1mm}
\section{Related Work}\label{sec:related_work}
\vspace{-2mm}

\textbf{Routing-based models }have been extensively studied to build modular language models~\citep{Cai2024ASO,switch_transformer}. MoE architectures consist of ``expert'' sub-models, each specialized for a particular domain or task, and a ``router'' that selects which expert to activate for a given input sequence or token. 
Multiple works have combined modular models and PEFT~\citep{ding2023parameter,hulora}. \citet{mixlora_li_2024} introduce Mix-LoRA for memory efficiency in language modeling;~\citet{mixlora_iclr} introduce Mix-LoRA to study domain specialization; and~\citet{mixlora_multimodal} introduce Mix-LoRA for multi-modal instruction tuning. Most similar to ours,~\citet{mixlora_domain} introduces Mix-LoRA for routing on domains like finance, medicine, and coding.
Other PEFT approaches include prefix-tuning~\citep{li2021prefix} or adapter learning~\citep{pmlr-v97-houlsby19a,hetowards}. We choose a hybrid of prompt-tuning~\citep{lester2021power}, which has been shown effective for privacy-aware fine-tuning~\citep{dp_prompt_pate}, and Mix-LoRA for effective domain experts.

\textbf{Preserving privacy }has mostly been studied with DP~\citep{dwork_dp}. However, DP approaches have primarily focused on applying DP to all model parameters~\citep{yu2021dpLLM,ghostclipping}. The benefits of leveraging public data during private fine-tuning were studied~\citep{kerrigan2020differentially,golatkar2022mixed}, showing that access to public data can improve the performance of DP-trained models. In our case, we use public data as the pre-trained model initialization and fine-tune with private datasets.
Similar to our setting,~\citet{tholoniat2024differentially} studies DP in an MoE setting but focuses on learned routers and emergent expertise, whereas we aim at private learning with positive transfer between domains.
Regarding the MIA, \citet{multimodal_MIA} explores a multi-modal attack, but our is about the multi-domain setting with access to predictive log-likelihoods.

\textbf{Knowledge Transfer in Modular Models }has been studied with large-scale pre-training with various data sources~\citep{radford2018improving,devlin_bert,touvron2023llama}. The term was introduced by~\citet{wmt_knowledge_transfer} and has been observed in the context of modular language models in the MultiCoder model~\citep{multicoder}. However, that work does not consider the privacy of each domain, nor does it study the privacy leakage in shared parameters of a neural network.

\vspace{-1mm}
\section{Conclusion}
\vspace{-2mm}

In this work, we have introduced a first-of-its-kind, hybrid PEFT architecture that we call Nexus-of-Experts. Our method, \tool, establishes a framework for modular learning that enables knowledge transfer across domains that contain private documents.
%
%
We apply our technique to the problem of multi-lingual code completion and our evaluation demonstrates that \tool\ effectively balances privacy and utility, achieving high accuracy across domains while providing strong guarantees at $(\varepsilon=1.,\delta=10^{-6})$-DP.
\tool bridges the accuracy gap between non-private and non-shared models by over 77\%.
At the same time, we uncover an empirical privacy leak in the baseline modular models and show how our method can defend against such attacks and protect training sources.


\bibliography{iclr2025_conference}

\begin{thebibliography}{67}
\providecommand{\natexlab}[1]{#1}
\providecommand{\url}[1]{\texttt{#1}}
\expandafter\ifx\csname urlstyle\endcsname\relax
  \providecommand{\doi}[1]{doi: #1}\else
  \providecommand{\doi}{doi: \begingroup \urlstyle{rm}\Url}\fi

\bibitem[Abadi et~al.(2016)Abadi, Chu, Goodfellow, McMahan, Mironov, Talwar, and Zhang]{abadi_dpsgd}
Martin Abadi, Andy Chu, Ian Goodfellow, H~Brendan McMahan, Ilya Mironov, Kunal Talwar, and Li~Zhang.
\newblock Deep learning with differential privacy.
\newblock In \emph{ACM SIGSAC conference on computer and communications security}, 2016.

\bibitem[Allamanis \& Sutton(2013)Allamanis and Sutton]{javagc}
Miltiadis Allamanis and Charles Sutton.
\newblock Mining source code repositories at massive scale using language modeling.
\newblock In \emph{Working conference on Mining Software Repositories (MSR)}. IEEE, 2013.

\bibitem[Bukaty(2019)]{ccpa}
Preston Bukaty.
\newblock \emph{The California Consumer Privacy Act (CCPA): An implementation guide}.
\newblock IT Governance Publishing, 2019.
\newblock ISBN 9781787781320.

\bibitem[Cai et~al.(2024)Cai, Jiang, Wang, Tang, Kim, and Huang]{Cai2024ASO}
Weilin Cai, Juyong Jiang, Fan Wang, Jing Tang, Sunghun Kim, and Jiayi Huang.
\newblock A survey on mixture of experts.
\newblock \emph{arXiv preprint 2407.06204}, 2024.

\bibitem[Carlini et~al.(2019)Carlini, Liu, Erlingsson, Kos, and Song]{carlini2019secret}
Nicholas Carlini, Chang Liu, {\'U}lfar Erlingsson, Jernej Kos, and Dawn Song.
\newblock The secret sharer: Evaluating and testing unintended memorization in neural networks.
\newblock In \emph{{USENIX} Security Symposium}, 2019.

\bibitem[Carlini et~al.(2021)Carlini, Tram{\`{e}}r, Wallace, Jagielski, Herbert{-}Voss, Lee, Roberts, Brown, Song, Erlingsson, Oprea, and Raffel]{carlini_ref_model_MIA}
Nicholas Carlini, Florian Tram{\`{e}}r, Eric Wallace, Matthew Jagielski, Ariel Herbert{-}Voss, Katherine Lee, Adam Roberts, Tom~B. Brown, Dawn Song, {\'{U}}lfar Erlingsson, Alina Oprea, and Colin Raffel.
\newblock Extracting training data from large language models.
\newblock In \emph{{USENIX} Security Symposium}, 2021.

\bibitem[Carlini et~al.(2022)Carlini, Chien, a~Milad~Nasr, Song, Terzis, and Tram{\`{e}}r]{MIA_argues_tpr}
Nicholas Carlini, Steve Chien, a~Milad~Nasr, Shuang Song, Andreas Terzis, and Florian Tram{\`{e}}r.
\newblock Membership inference attacks from first principles.
\newblock In \emph{{IEEE} Symposium on Security and Privacy, {SP}}, 2022.

\bibitem[Chang et~al.(2024)Chang, Shamsabadi, Katevas, Haddadi, and Shokri]{camia}
Hongyan Chang, Ali~Shahin Shamsabadi, Kleomenis Katevas, Hamed Haddadi, and Reza Shokri.
\newblock Context-aware membership inference attacks against pre-trained large language models.
\newblock \emph{arXiv preprint 2409.13745}, 2024.

\bibitem[{Chen~et~al.}(2021)]{human_eval}
{Chen~et~al.}
\newblock Evaluating large language models trained on code.
\newblock \emph{arXiv preprint 2107.03374}, 2021.

\bibitem[CodeParrot(2025)]{codeparrot_github_code}
CodeParrot.
\newblock Github code dataset, 2025.
\newblock URL \url{https://huggingface.co/datasets/codeparrot/github-code}.
\newblock Accessed: 2025-01-09.

\bibitem[Devlin et~al.(2019)Devlin, Chang, Lee, and Toutanova]{devlin_bert}
Jacob Devlin, Ming{-}Wei Chang, Kenton Lee, and Kristina Toutanova.
\newblock {BERT:} pre-training of deep bidirectional transformers for language understanding.
\newblock In \emph{Proceedings of the North American Chapter of the Association for Computational Linguistics: Human Language Technologies, {NAACL-HLT}}. Association for Computational Linguistics, 2019.

\bibitem[Ding et~al.(2023)Ding, Qin, Yang, Wei, Yang, Su, Hu, Chen, Chan, Chen, et~al.]{ding2023parameter}
Ning Ding, Yujia Qin, Guang Yang, Fuchao Wei, Zonghan Yang, Yusheng Su, Shengding Hu, Yulin Chen, Chi-Min Chan, Weize Chen, et~al.
\newblock Parameter-efficient fine-tuning of large-scale pre-trained language models.
\newblock \emph{Nature Machine Intelligence}, 2023.

\bibitem[Dong et~al.(2023)Dong, Moon, Xu, Malik, and Yu]{dong2023towards}
Xin~Luna Dong, Seungwhan Moon, Yifan~Ethan Xu, Kshitiz Malik, and Zhou Yu.
\newblock Towards next-generation intelligent assistants leveraging llm techniques.
\newblock In \emph{Proceedings of the SIGKDD Conference on Knowledge Discovery and Data Mining}, 2023.

\bibitem[Duan et~al.(2023)Duan, Dziedzic, Papernot, and Boenisch]{dp_prompt_pate}
Haonan Duan, Adam Dziedzic, Nicolas Papernot, and Franziska Boenisch.
\newblock Flocks of stochastic parrots: Differentially private prompt learning for large language models.
\newblock In \emph{Advances in Neural Information Processing Systems (NeurIPS)}, 2023.

\bibitem[Dwork \& Roth(2014)Dwork and Roth]{dwork_dp}
Cynthia Dwork and Aaron Roth.
\newblock The algorithmic foundations of differential privacy.
\newblock \emph{Foundations and Trends in Theoretical Computer Science}, 2014.

\bibitem[EU-regulation(2024)]{eurlex2024}
EU-regulation.
\newblock Regulation (eu) 2024/1689, 2024.
\newblock URL \url{https://eur-lex.europa.eu/legal-content/EN/TXT/?uri=CELEX%3A32024R1689}.
\newblock Accessed: 2024-11-11.

\bibitem[Fedus et~al.(2022)Fedus, Zoph, and Shazeer]{switch_transformer}
William Fedus, Barret Zoph, and Noam Shazeer.
\newblock Switch transformers: Scaling to trillion parameter models with simple and efficient sparsity.
\newblock \emph{Journal of Machine Learning Research}, 2022.

\bibitem[Feng et~al.(2024)Feng, Hao, Zhang, Han, and Wang]{mixlora_domain}
Wenfeng Feng, Chuzhan Hao, Yuewei Zhang, Yu~Han, and Hao Wang.
\newblock Mixture-of-loras: An efficient multitask tuning method for large language models.
\newblock In \emph{Proceedings of the Joint International Conference on Computational Linguistics, Language Resources and Evaluation, {LREC/COLING}}. {ELRA} and {ICCL}, 2024.

\bibitem[{Foo Yun Chee}(2024)]{apple_eu}
{Foo Yun Chee}.
\newblock Apple to delay launch of ai-powered features in europe, blames eu tech rules, 2024.
\newblock URL \url{https://www.reuters.com/technology/artificial-intelligence/ apple-delay-launch-ai-powered\ -features-europe-blames-eu-tech-rules\ -2024-06-21/}.
\newblock Accessed: 2024-12-05.

\bibitem[GitHub(2024)]{githubcopilot}
GitHub.
\newblock Github copilot, 2024.
\newblock URL \url{https://github.com/features/copilot}.
\newblock Accessed: 2024-11-11.

\bibitem[Golatkar et~al.(2022)Golatkar, Achille, Wang, Roth, Kearns, and Soatto]{golatkar2022mixed}
Aditya Golatkar, Alessandro Achille, Yu-Xiang Wang, Aaron Roth, Michael Kearns, and Stefano Soatto.
\newblock Mixed differential privacy in computer vision.
\newblock In \emph{Proceedings of the Conference on Computer Vision and Pattern Recognition (CVPR)}, 2022.

\bibitem[Gong et~al.(2022)Gong, Guo, Zhou, Gao, Wang, and Xu]{multicoder}
Zi~Gong, Yinpeng Guo, Pingyi Zhou, Cuiyun Gao, Yasheng Wang, and Zenglin Xu.
\newblock Multicoder: Multi-programming-lingual pre-training for low-resource code completion.
\newblock \emph{arXiv preprint 2212.09666}, 2022.

\bibitem[Gururangan et~al.(2022)Gururangan, Lewis, Holtzman, Smith, and Zettlemoyer]{gururangan-etal-2022-demix}
Suchin Gururangan, Mike Lewis, Ari Holtzman, Noah~A. Smith, and Luke Zettlemoyer.
\newblock {DEM}ix layers: Disentangling domains for modular language modeling.
\newblock In \emph{Proceedings of the Annual Meeting of the Association for Computational Linguistics (ACL)}, 2022.

\bibitem[He et~al.(2021)He, Zhou, Ma, Berg-Kirkpatrick, and Neubig]{hetowards}
Junxian He, Chunting Zhou, Xuezhe Ma, Taylor Berg-Kirkpatrick, and Graham Neubig.
\newblock Towards a unified view of parameter-efficient transfer learning.
\newblock In \emph{International Conference on Learning Representations (ICLR)}, 2021.

\bibitem[Hokamp et~al.(2019)Hokamp, Glover, and Ghalandari]{wmt_knowledge_transfer}
Chris Hokamp, John Glover, and Demian~Gholipour Ghalandari.
\newblock Evaluating the supervised and zero-shot performance of multi-lingual translation models.
\newblock In \emph{Conference on Machine Translation, {WMT}}. Association for Computational Linguistics, 2019.

\bibitem[Houlsby et~al.(2019)Houlsby, Giurgiu, Jastrzebski, Morrone, De~Laroussilhe, Gesmundo, Attariyan, and Gelly]{pmlr-v97-houlsby19a}
Neil Houlsby, Andrei Giurgiu, Stanislaw Jastrzebski, Bruna Morrone, Quentin De~Laroussilhe, Andrea Gesmundo, Mona Attariyan, and Sylvain Gelly.
\newblock Parameter-efficient transfer learning for {NLP}.
\newblock In \emph{International Conference on Machine Learning (ICML)}, Proceedings of Machine Learning Research. PMLR, 2019.

\bibitem[Hsu et~al.(2014)Hsu, Gaboardi, Haeberlen, Khanna, Narayan, Pierce, and Roth]{choosing_epsilon}
Justin Hsu, Marco Gaboardi, Andreas Haeberlen, Sanjeev Khanna, Arjun Narayan, Benjamin~C Pierce, and Aaron Roth.
\newblock Differential privacy: An economic method for choosing epsilon.
\newblock In \emph{IEEE Computer Security Foundations Symposium}. IEEE, 2014.

\bibitem[Hu et~al.(2021)Hu, Wallis, Allen-Zhu, Li, Wang, Wang, Chen, et~al.]{hulora}
Edward~J Hu, Phillip Wallis, Zeyuan Allen-Zhu, Yuanzhi Li, Shean Wang, Lu~Wang, Weizhu Chen, et~al.
\newblock Lora: Low-rank adaptation of large language models.
\newblock In \emph{International Conference on Learning Representations (ICLR)}, 2021.

\bibitem[Hu et~al.(2022)Hu, Wang, Sun, Wang, and Xue]{multimodal_MIA}
Pingyi Hu, Zihan Wang, Ruoxi Sun, Hu~Wang, and Minhui Xue.
\newblock M{\textdollar}{\^{}}4{\textdollar}i: Multi-modal models membership inference.
\newblock In \emph{Advances in Neural Information Processing Systems (NeurIPS)}, 2022.

\bibitem[Husain et~al.(2019)Husain, Wu, Gazit, Allamanis, and Brockschmidt]{codesearchnet}
Hamel Husain, Ho{-}Hsiang Wu, Tiferet Gazit, Miltiadis Allamanis, and Marc Brockschmidt.
\newblock Codesearchnet challenge: Evaluating the state of semantic code search.
\newblock \emph{arXiv preprint 1909.09436}, 2019.

\bibitem[Iacob et~al.(2024)Iacob, Sani, Marino, Aleksandrov, Shen, and Lane]{iacob2024worldwide}
Alex Iacob, Lorenzo Sani, Bill Marino, Preslav Aleksandrov, William~F Shen, and Nicholas~Donald Lane.
\newblock Worldwide federated training of language models.
\newblock \emph{arXiv preprint 2405.14446}, 2024.

\bibitem[Kaplan et~al.(2020)Kaplan, McCandlish, Henighan, Brown, Chess, Child, Gray, Radford, Wu, and Amodei]{kaplan2020scaling}
Jared Kaplan, Sam McCandlish, Tom Henighan, Tom~B Brown, Benjamin Chess, Rewon Child, Scott Gray, Alec Radford, Jeffrey Wu, and Dario Amodei.
\newblock Scaling laws for neural language models.
\newblock \emph{arXiv preprint 2001.08361}, 2020.

\bibitem[Kerrigan et~al.(2020)Kerrigan, Slack, and Tuyls]{kerrigan2020differentially}
Gavin Kerrigan, Dylan Slack, and Jens Tuyls.
\newblock Differentially private language models benefit from public pre-training.
\newblock In \emph{Proceedings of the Second Workshop on Privacy in NLP}, 2020.

\bibitem[Komatsuzaki et~al.(2022)Komatsuzaki, Puigcerver, Lee-Thorp, Ruiz, Mustafa, Ainslie, Tay, Dehghani, and Houlsby]{komatsuzaki2022sparse}
Aran Komatsuzaki, Joan Puigcerver, James Lee-Thorp, Carlos~Riquelme Ruiz, Basil Mustafa, Joshua Ainslie, Yi~Tay, Mostafa Dehghani, and Neil Houlsby.
\newblock Sparse upcycling: Training mixture-of-experts from dense checkpoints.
\newblock \emph{arXiv preprint 2212.05055}, 2022.

\bibitem[Kopiczko et~al.(2024)Kopiczko, Blankevoort, and Asano]{kopiczkovera}
Dawid~Jan Kopiczko, Tijmen Blankevoort, and Yuki~M Asano.
\newblock Vera: Vector-based random matrix adaptation.
\newblock In \emph{International Conference on Learning Representations (ICLR)}, 2024.

\bibitem[Kotalwar et~al.(2024)Kotalwar, Gotovos, and Singla]{kotalwar2024hints}
Nachiket Kotalwar, Alkis Gotovos, and Adish Singla.
\newblock Hints-in-browser: Benchmarking language models for programming feedback generation.
\newblock \emph{arXiv preprint 2406.05053}, 2024.

\bibitem[Lepikhin et~al.(2020)Lepikhin, Lee, Xu, Chen, Firat, Huang, Krikun, Shazeer, and Chen]{gshard}
Dmitry Lepikhin, HyoukJoong Lee, Yuanzhong Xu, Dehao Chen, Orhan Firat, Yanping Huang, Maxim Krikun, Noam Shazeer, and Zhifeng Chen.
\newblock Gshard: Scaling giant models with conditional computation and automatic sharding.
\newblock \emph{arXiv preprint 2006.16668}, 2020.

\bibitem[Lester et~al.(2021)Lester, Al{-}Rfou, and Constant]{lester2021power}
Brian Lester, Rami Al{-}Rfou, and Noah Constant.
\newblock The power of scale for parameter-efficient prompt tuning.
\newblock In \emph{Proceedings of the Conference on Empirical Methods in Natural Language Processing, (EMNLP)}. Association for Computational Linguistics, 2021.

\bibitem[Li et~al.(2024)Li, Ma, Wang, Cheng, Duan, Zuo, Yang, and Tang]{mixlora_li_2024}
Dengchun Li, Yingzi Ma, Naizheng Wang, Zhiyuan Cheng, Lei Duan, Jie Zuo, Cal Yang, and Mingjie Tang.
\newblock Mixlora: Enhancing large language models fine-tuning with lora based mixture of experts.
\newblock \emph{arXiv preprint 2404.15159}, 2024.

\bibitem[Li \& Liang(2021)Li and Liang]{li2021prefix}
Xiang~Lisa Li and Percy Liang.
\newblock Prefix-tuning: Optimizing continuous prompts for generation.
\newblock In \emph{Proceedings of the Annual Meeting of the Association for Computational Linguistics (ACL)}. Association for Computational Linguistics, 2021.

\bibitem[Li et~al.(2021)Li, Tramer, Liang, and Hashimoto]{ghostclipping}
Xuechen Li, Florian Tramer, Percy Liang, and Tatsunori Hashimoto.
\newblock Large language models can be strong differentially private learners.
\newblock In \emph{International Conference on Learning Representations (ICLR)}, 2021.

\bibitem[Liu et~al.(2024)Liu, Wang, Yin, Molchanov, Wang, Cheng, and Chen]{liudora}
Shih-yang Liu, Chien-Yi Wang, Hongxu Yin, Pavlo Molchanov, Yu-Chiang~Frank Wang, Kwang-Ting Cheng, and Min-Hung Chen.
\newblock Dora: Weight-decomposed low-rank adaptation.
\newblock In \emph{International Conference on Machine Learning (ICML)}, 2024.

\bibitem[Loshchilov \& Hutter(2019)Loshchilov and Hutter]{AdamW}
Ilya Loshchilov and Frank Hutter.
\newblock Decoupled weight decay regularization.
\newblock In \emph{International Conference on Learning Representations (ICLR)}, 2019.

\bibitem[Lu et~al.(2021)Lu, Guo, Ren, Huang, Svyatkovskiy, Blanco, Clement, Drain, Jiang, Tang, et~al.]{lu2021codexglue}
Shuai Lu, Daya Guo, Shuo Ren, Junjie Huang, Alexey Svyatkovskiy, Ambrosio Blanco, Colin Clement, Dawn Drain, Daxin Jiang, Duyu Tang, et~al.
\newblock Codexglue: A machine learning benchmark dataset for code understanding and generation.
\newblock \emph{arXiv preprint 2102.04664}, 2021.

\bibitem[Niu et~al.(2023)Niu, Mirza, Maradni, and P{\"o}pper]{291327}
Liang Niu, Shujaat Mirza, Zayd Maradni, and Christina P{\"o}pper.
\newblock {CodexLeaks}: Privacy leaks from code generation language models in {GitHub} copilot.
\newblock In \emph{{USENIX} Security Symposium}, 2023.

\bibitem[{OpenAI}(2023)]{openai2023gpt35}
{OpenAI}.
\newblock Gpt-3.5 turbo fine-tuning and api updates, August 2023.
\newblock URL \url{https://openai.com/index/gpt-3-5-turbo-fine-tuning-and-api-updates/}.
\newblock Accessed: 2025-01-26.

\bibitem[Pedregosa et~al.(2011)Pedregosa, Varoquaux, Gramfort, Michel, Thirion, Grisel, Blondel, Prettenhofer, Weiss, Dubourg, Vanderplas, Passos, Cournapeau, Brucher, Perrot, and Duchesnay]{sklearn}
F.~Pedregosa, G.~Varoquaux, A.~Gramfort, V.~Michel, B.~Thirion, O.~Grisel, M.~Blondel, P.~Prettenhofer, R.~Weiss, V.~Dubourg, J.~Vanderplas, A.~Passos, D.~Cournapeau, M.~Brucher, M.~Perrot, and E.~Duchesnay.
\newblock Scikit-learn: Machine learning in {P}ython.
\newblock \emph{Journal of Machine Learning Research}, 2011.

\bibitem[Pfeiffer et~al.(2023)Pfeiffer, Ruder, Vuli{\'c}, and Ponti]{pfeiffer2023modular}
Jonas Pfeiffer, Sebastian Ruder, Ivan Vuli{\'c}, and Edoardo Ponti.
\newblock Modular deep learning.
\newblock \emph{Transactions on Machine Learning Research}, 2023.

\bibitem[Radford(2018)]{radford2018improving}
Alec Radford.
\newblock Improving language understanding by generative pre-training.
\newblock \emph{OpenAI distribution network}, 2018.

\bibitem[Raychev et~al.(2016)Raychev, Bielik, and Vechev]{py150}
Veselin Raychev, Pavol Bielik, and Martin Vechev.
\newblock Probabilistic model for code with decision trees.
\newblock \emph{ACM SIGPLAN Notices}, 2016.

\bibitem[Ruder et~al.(2019)Ruder, Peters, Swayamdipta, and Wolf]{ruder-etal-2019-transfer}
Sebastian Ruder, Matthew~E. Peters, Swabha Swayamdipta, and Thomas Wolf.
\newblock Transfer learning in natural language processing.
\newblock In Anoop Sarkar and Michael Strube (eds.), \emph{Proceedings of the Annual Meeting of the Association for Computational Linguistics (ACL)}, Minneapolis, Minnesota, 2019.

\bibitem[{Salesforce}(2024)]{salesforce2024agentforce}
{Salesforce}.
\newblock Introducing agentforce 2.0: The digital labor platform for building a limitless workforce, December 2024.
\newblock URL \url{https://www.salesforce.com/uk/news/press-releases/2024/12/17/agentforce-2-0-announcement/}.
\newblock Accessed: 2025-01-26.

\bibitem[Shen et~al.(2024)Shen, Xu, Wang, Cheng, Yin, and Huang]{mixlora_multimodal}
Ying Shen, Zhiyang Xu, Qifan Wang, Yu~Cheng, Wenpeng Yin, and Lifu Huang.
\newblock Multimodal instruction tuning with conditional mixture of lora.
\newblock In \emph{Proceedings of the Annual Meeting of the Association for Computational Linguistics (ACL)}. Association for Computational Linguistics, 2024.

\bibitem[Shokri et~al.(2017)Shokri, Stronati, Song, and Shmatikov]{shokri2017membership}
Reza Shokri, Marco Stronati, Congzheng Song, and Vitaly Shmatikov.
\newblock Membership inference attacks against machine learning models.
\newblock In \emph{Symposium on security and privacy (SP)}. IEEE, 2017.

\bibitem[Shumailov et~al.(2024)Shumailov, Shumaylov, Zhao, Papernot, Anderson, and Gal]{shumailov2024ai}
Ilia Shumailov, Zakhar Shumaylov, Yiren Zhao, Nicolas Papernot, Ross Anderson, and Yarin Gal.
\newblock {AI} models collapse when trained on recursively generated data.
\newblock \emph{Nature}, 631, 2024.

\bibitem[{Stefano Fratta}(2024)]{meta_eu}
{Stefano Fratta}.
\newblock Building ai technology for europeans in a transparent and responsible way, 2024.
\newblock URL \url{https://about.fb.com/news/2024/06/ building-ai-technology- for-europeans \ -in-a-transparent-and-responsible-way/}.
\newblock Accessed: 2024-12-05.

\bibitem[Tholoniat et~al.(2024)Tholoniat, Inan, Kulkarni, and Sim]{tholoniat2024differentially}
Pierre Tholoniat, Huseyin~A Inan, Janardhan Kulkarni, and Robert Sim.
\newblock Differentially private training of mixture of experts models.
\newblock \emph{AAAI privacy workshop}, 2024.

\bibitem[Touvron et~al.(2023)Touvron, Lavril, Izacard, Martinet, Lachaux, Lacroix, Rozi{\`e}re, Goyal, Hambro, Azhar, et~al.]{touvron2023llama}
Hugo Touvron, Thibaut Lavril, Gautier Izacard, Xavier Martinet, Marie-Anne Lachaux, Timoth{\'e}e Lacroix, Baptiste Rozi{\`e}re, Naman Goyal, Eric Hambro, Faisal Azhar, et~al.
\newblock Llama: Open and efficient foundation language models.
\newblock \emph{arXiv preprint 2302.13971}, 2023.

\bibitem[Wang et~al.(2023)Wang, Le, Gotmare, Bui, Li, and Hoi]{codet5p}
Yue Wang, Hung Le, Akhilesh Gotmare, Nghi D.~Q. Bui, Junnan Li, and Steven C.~H. Hoi.
\newblock Codet5+: Open code large language models for code understanding and generation.
\newblock In \emph{Proceedings of the Conference on Empirical Methods in Natural Language Processing, (EMNLP)}, 2023.

\bibitem[Wu et~al.(2024)Wu, Huang, and Wei]{mixlora_iclr}
Xun Wu, Shaohan Huang, and Furu Wei.
\newblock Mixture of lora experts.
\newblock In \emph{International Conference on Learning Representations (ICLR)}, 2024.

\bibitem[Xu et~al.(2024)Xu, Li, Xu, and Zhang]{xu-etal-2024-llms}
Jialiang Xu, Shenglan Li, Zhaozhuo Xu, and Denghui Zhang.
\newblock Do {LLM}s know to respect copyright notice?
\newblock In Yaser Al-Onaizan, Mohit Bansal, and Yun-Nung Chen (eds.), \emph{Proceedings of the Conference on Empirical Methods in Natural Language Processing, (EMNLP)}. Association for Computational Linguistics, 2024.

\bibitem[Yousefpour et~al.(2021)Yousefpour, Shilov, Sablayrolles, Testuggine, Prasad, Malek, Nguyen, Ghosh, Bharadwaj, Zhao, Cormode, and Mironov]{opacus}
Ashkan Yousefpour, Igor Shilov, Alexandre Sablayrolles, Davide Testuggine, Karthik Prasad, Mani Malek, John Nguyen, Sayan Ghosh, Akash Bharadwaj, Jessica Zhao, Graham Cormode, and Ilya Mironov.
\newblock Opacus: User-friendly differential privacy library in pytorch.
\newblock \emph{arXiv preprint 2109.12298}, 2021.

\bibitem[Yu et~al.(2021{\natexlab{a}})Yu, Naik, Backurs, Gopi, Inan, Kamath, Kulkarni, Lee, Manoel, Wutschitz, et~al.]{yu2021dpLLM}
Da~Yu, Saurabh Naik, Arturs Backurs, Sivakanth Gopi, Huseyin~A Inan, Gautam Kamath, Janardhan Kulkarni, Yin~Tat Lee, Andre Manoel, Lukas Wutschitz, et~al.
\newblock Differentially private fine-tuning of language models.
\newblock In \emph{International Conference on Learning Representations (ICLR)}, 2021{\natexlab{a}}.

\bibitem[Yu et~al.(2021{\natexlab{b}})Yu, Zhang, Chen, Yin, and Liu]{yu2021large}
Da~Yu, Huishuai Zhang, Wei Chen, Jian Yin, and Tie-Yan Liu.
\newblock Large scale private learning via low-rank reparametrization.
\newblock In \emph{International Conference on Machine Learning (ICML)}. PMLR, 2021{\natexlab{b}}.

\bibitem[Zheng et~al.(2024)Zheng, Gou, Kil, Sun, and Su]{zhenggpt}
Boyuan Zheng, Boyu Gou, Jihyung Kil, Huan Sun, and Yu~Su.
\newblock Gpt-4v (ision) is a generalist web agent, if grounded.
\newblock In \emph{International Conference on Machine Learning (ICML)}, 2024.

\bibitem[Zhou et~al.(2024)Zhou, Wang, Huang, Huang, Han, Feng, Deng, Luo, and Chen]{zhou2024moe}
Hao Zhou, Zhijun Wang, Shujian Huang, Xin Huang, Xue Han, Junlan Feng, Chao Deng, Weihua Luo, and Jiajun Chen.
\newblock Moe-lpr: Multilingual extension of large language models through mixture-of-experts with language priors routing.
\newblock \emph{arXiv preprint 2408.11396}, 2024.

\bibitem[Zou et~al.(2023)Zou, Wang, Carlini, Nasr, Kolter, and Fredrikson]{zou2023universal}
Andy Zou, Zifan Wang, Nicholas Carlini, Milad Nasr, J~Zico Kolter, and Matt Fredrikson.
\newblock Universal and transferable adversarial attacks on aligned language models.
\newblock \emph{arXiv preprint 2307.15043}, 2023.

\end{thebibliography}
\bibliographystyle{iclr2025_conference}

\appendix

\newpage
\section{Limitations \& Future Work}

We have laid the foundations for a new, privacy-friendly, modular method of addressing domain privacy in LLMs. However, this is an important problem with many additional dimensions to consider.

The approach has been tested against a relatively small LLM, i.e.,~\codetp (220M parameters) and a single dataset (\codexglue). While enough for making a case towards this new paradigm, we aim to extend our evaluation to additional domains and tasks, such as use-cases in Natural Language.
Moreover, while largely orthogonal, we have leveraged LoRAs as our PEFT technique of choice for domain experts. Nevertheless, there have been various alternative architectures for adaptation, including but not limited to DoRA~\citep{liudora} and VeRA~\citep{kopiczkovera}. We aim to explore such alternative forms for parameter-efficient adapters and their interplay with NoE in future work.
There is also the question of achieving modularity, privacy, and knowledge transfer in a single-stage training method, instead of the two stages in Algorithm~\ref{alg:mixed_priv_dpsgd}, but that must ensure that gradients from private and locally non-private samples do not interact. 
Finally, we uncover and defend against a single type of attack, namely the Membership Inference Attack. However, there are different possible threat models and attacks (e.g., adversarial~\citep{zou2023universal} and data reconstruction~\citep{carlini2019secret}). While not in the direct scope of our study, we do acknowledge their importance in various deployment settings and future avenues of work.

\section{Ablation Study} \label{sec:ablation}
\begin{table}[h]

\end{table}

\begin{wraptable}{lh}{8cm}
\centering
\vspace{-4mm}
\setlength{\tabcolsep}{4pt}
\caption{Ablation: Surgically removing the shared parameters, after the training has finished, leads to the most accuracy decrease in the domain with the most data, Python. Removing the domain experts leads to the most accuracy decrease in the domain with the least data, Go.}
\begin{tabular}{l | c | r r r}
\toprule
\textbf{Model} & \textbf{$\varepsilon$}  & \textbf{Python} & \textbf{Java} & \textbf{Go} \\
\midrule
\tool\ (pt32) & 1.0 & 69.14 & 61.19 & 66.58 \\ 
\midrule
w/out common prompts & 0.0 & 4.65 & 14.31 & 15.42 \\
w/out domain experts & 1.0 & 33.91 & 24.30 & 19.13 \\
\bottomrule
\end{tabular}
\label{tab:ablation}
\vspace{-0.2cm}
\end{wraptable}

We perform an ablation study to understand the impact of each component of \tool\ on the final performance. In this setting, we surgically remove the shared parameters to investigate the impact on the results. The results are shown in Table~\ref{tab:ablation}.
First off, we observe that removing any of the components of \tool is detrimental to its downstream accuracy, with a minimum drop of 36.89\% points and an average of 50\% for common prompts and 30\% for domain experts. The common prompt seems to be more important in the downstream accuracy, but this can be an effect of our training stage ordering. However, it yields an important conclusion that LoRA experts do not learn or ``undo'' the knowledge of prompt tokens, but incorporate it into their knowledge. Moreover, the common prompts are not a nuisance and do encode important bits of information across domains. Finally, and in line with the comparison with the Share-Nothing (\textit{(i)}), it seems that the positive transfer (whether by removal of common prompts or by fine-tuning without them) more severely affects smaller datasets and especially Go with $50\times$ fewer documents than Python. Therefore, scarce datasets tend to benefit more from knowledge sharing. 

For the analysis in Fig.~\ref{fig:bridge}, the results for the scarce language, Go, are particularly noteworthy. The knowledge transfer is more than 2.4\% points over the Shared Nothing setup. \emph{This highlights the effectiveness of our approach in leveraging shared knowledge across domains to enhance token completion tasks, especially for a lower-resource domain.}

\section{Evaluation Setup Details}
\begin{table}[h]
\centering
\caption{An overview of datasets and domains used for the training and evaluation of \tool\ on code completion tasks.}
\vspace{1mm}
\setlength{\tabcolsep}{4pt}
\begin{adjustbox}{max width=\textwidth}
\begin{tabular}{l l l l l}
\toprule
\textbf{Domain} & \textbf{Source} & \textbf{Description} & \textbf{Training Set} & \textbf{Evaluation Set}  \\
\midrule
Pre-training &  GitHub Code~\citep{codeparrot_github_code} & Upstream data & \multicolumn{2}{c}{3.7 million }  \\  
\midrule
Python & PY150~\citep{py150} & Published dataset$^\dagger$ & \num{100000}  & \num{50000} \\
Java & Java GitHub Corpus~\citep{javagc} & Published dataset$^\dagger$ & \num{12934}  & \num{8268} \\
Go & \codesearchnet~\citep{codesearchnet}  & Extracted from \codesearchnet\ $^\dagger$ & \num{2000}  & \num{1559} \\
\bottomrule
\multicolumn{4}{l}{$^\dagger$ Dataset explicitly deduplicated in pre-training dataset~\citep{codet5p}.} \\
\end{tabular}
\end{adjustbox}
\vspace{-0.2cm}
\label{tab:datasets}
\end{table}

\subsection{Dataset Details}  \label{app:dataset}

This section provides additional details about the datasets we use for training, their partitioning, and deduplication against the pre-training set of \codesearchnet.

\textbf{Python Dataset. } The \pyonefifty\ data set consists of \num{150000} Python repositories~\citep{py150}. Each ``document'' is the code from a GitHub repository collected in 2016. The original publishers of the data set did select for repositories with permissive licenses. While originally created for code parsing, we use only the actual code tokens for next-token prediction. 

\textbf{Java Dataset.} The \javacorpus\ data set consists of \num{14807} Java code repositories~\citep{javagc}. Each ``document'' is the code from a GitHub repository collected in 2013. The original publishers of the data set did select for repositories with permissive licences and repositories that were forked at least once. Additionally, repositories with a ``below average'' reputation score on GitHub are not considered. In 2021, the \codexglue\ effort repeated this strategy to collect another \num{6395} code repositories under the same collection policy~\citep{lu2021codexglue}.

\textbf{Go dataset.} For Go, we make a new train-test split from the \codesearchnet\ dataset, which we know was also deduplicated against by~\citep{codet5p}. As \codesearchnet\ has data for a text-to-code search task, we remove the search queries and concatenate all functions of a repository to obtain a document. The data set sizes are noted in~\Cref{tab:datasets}.

The data set for Go consists of \num{167288} search queries originating from \num{3559} repositories. We split the data set into \num{2000} repositories for training and \num{1559} repositories for testing. Functions are concatenated with interleaved linebreaks. This mimics the provisioning strategy for the \pyonefifty\ and \javacorpus\ datasets, where each document represents a complete code library and thus contains multiple functions.

\begin{figure}[h]
\centering
\vspace{3mm}
\includegraphics[width=0.9\linewidth]{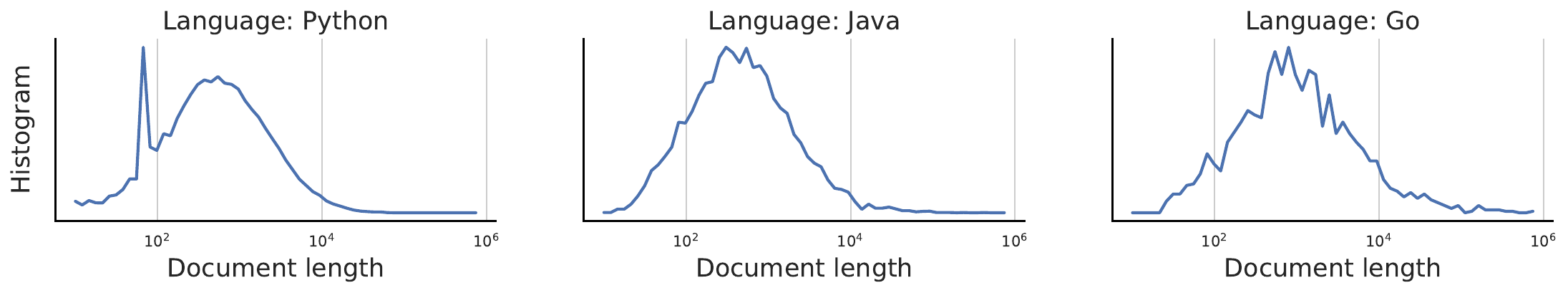}
\caption{Histograms of number of tokens per document for the three domains. The distributions are similar in order of magnitude. The dataset sizes are different: Python has \num{100000} documents for training, Java \num{12934}, and Go \num{2000}. }
\label{fig:dataset_histogram}
\end{figure}

\textbf{Histograms: } Fig.~\ref{fig:dataset_histogram} shows histograms of the number of tokens per document for the training sets of the three domains. This indicates that the described provisioning method for Go has a similar order of magnitude of tokens per document as the Python and Java datasets. The histogram for Go is more noisy since the dataset contains only \num{2000} documents, compared to \num{100000} documents for Python. The histogram for Python shows a single peak at 81 tokens per document, which is a boilerplate script that is repeated several times in the training set.

We use the data sets as they are tokenized by \codexglue. Long tokens might be split up by the \codetp\ tokenizer, similar to the training algorithm of the pre-trained model~\citep{codet5p}. Examples of tokens as blank-spaced strings are in Section~\ref{app:examples}.

\subsection{Baseline Details}
\label{app:baselines}

\begin{figure*}[h]
\centering
\begin{subfigure}[t]{0.4\textwidth}
    \centering
    \includegraphics[width=\linewidth,trim={0.6cm 0.3cm 2cm 0.4cm},clip]{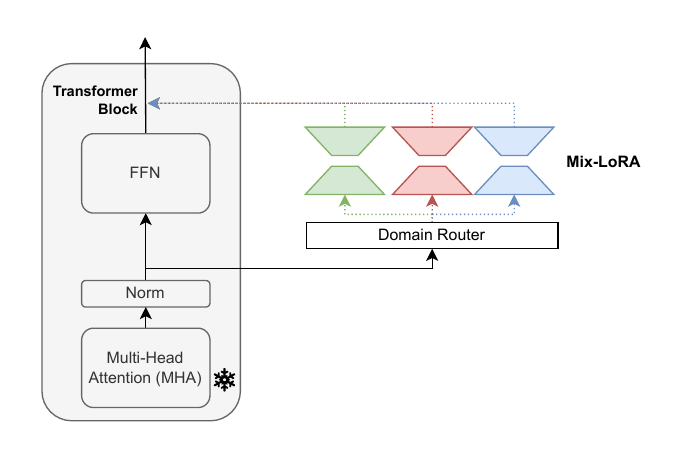}
    \caption{Share Nothing}
    \label{fig:share-nothing}
\end{subfigure}
\begin{subfigure}[t]{0.19\textwidth}
    \centering
    \includegraphics[width=\linewidth,trim={0.6cm 0.8cm 0.6cm 0},clip]{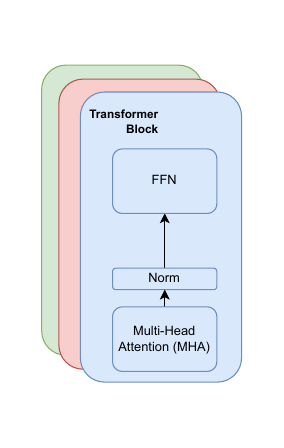}
    \caption{Solo}
    \label{fig:solo}
\end{subfigure}
\begin{subfigure}[t]{0.17\textwidth}
    \centering
    \includegraphics[width=\linewidth,trim={0.6cm 1.2cm 0.6cm 0},clip]{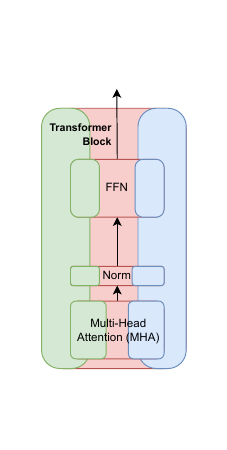}
    \caption{Monolithic Fine-tuning}
    \label{fig:monolithic-finetuning}
\end{subfigure} \\ 
\begin{subfigure}[t]{0.29\textwidth}
    \centering
    \includegraphics[width=\linewidth,trim={0.6cm 0.3cm 0.6cm 0.75cm},clip]{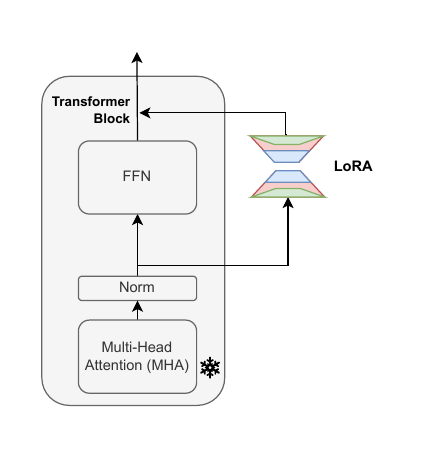}
    \caption{Single Common Adapter}
    \label{fig:single-common-adapter}
\end{subfigure}
\begin{subfigure}[t]{0.15\textwidth}
    \centering
    \includegraphics[width=\linewidth,trim={0.6cm 1.2cm 0.6cm 0.75cm},clip]{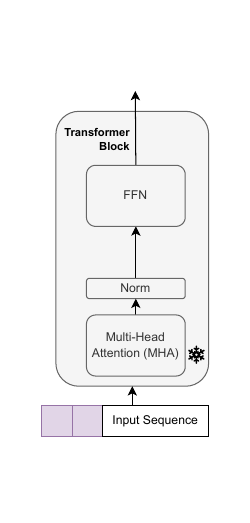}
    \caption{Prompt-Tuning Only}
    \label{fig:prompt-tuning-only}
\end{subfigure}
\begin{subfigure}[t]{0.43\textwidth}
    \centering
    \includegraphics[width=\linewidth,trim={0.6cm 0.3cm 0.6cm 0.65cm},clip]{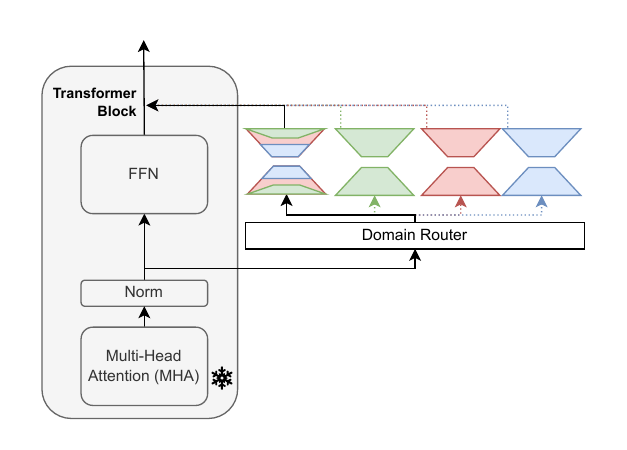} 
    \caption{\tool with Common LoRA}
    \label{fig:noesis-with-common-lora}
\end{subfigure}

\caption{Illustration of different baselines. Grey represents frozen, while blue, green, and red represent the domains with which we train that specific component. Purple represents trainable prompts.}
\label{fig:enter-label}
\end{figure*}

In this section, we describe each baseline in greater detail and why we compare to it. Specifically:

\begin{itemize}[label={},left=0pt]
\item \textbf{\textit{i)}~Share Nothing} (Fig.~\ref{fig:share-nothing}): This setup represents the architecture where we train domain adapters of a model without any common parameters. The backbone model remains frozen. This effectively represents a setup which can have no knowledge transfer between the individual domains. In other words, there is no parameter whose gradient depends on more than one domain.
\item \textbf{\textit{ii)}~Solo} (Fig.~\ref{fig:solo}): Solo represents a set of models, each trained with DP-SGD individually on a specific domain. In this case, the baseline only shares the same pre-training between the individual models and diverges during private fine-tuning.
\item \textbf{\textit{iii)}~Monolithic Fine-tuning} (Fig.~\ref{fig:monolithic-finetuning}): This setup represents the full fine-tuning of a single model on all datasets jointly. It exemplifies the paradigm with the largest potential for positive transfer but is also prone to overfitting as the setup has many trainable parameters.
\item \textbf{\textit{iv)}~Single Common Adapter} (Fig.~\ref{fig:single-common-adapter}): This setup represents the LoRA variant of \textit{(iii)}, where we tune a common LoRA with rank four across all domains. This baseline is to corroborate the increase in accuracy when using PEFT in a training algorithm with DP guarantees~\citep{yu2021dpLLM,yu2021large}.
\item \textbf{\textit{v)}~Prompt-Tuning Only}(Fig.~\ref{fig:prompt-tuning-only}): This baseline represents the training of 32 common prompt tokens across all domains only, with the rest of the backbone being frozen. It represents the most lightweight PEFT variant here.
\end{itemize}

\subsubsection{\tool with LoRA as Knowledge Sharing Backbone}
\label{app:noesis_rc}

Table~\ref{tab:common_params} compares \tool\ against a variant with the ``shared parameters'' in the form of another LoRA~\citep{mixlora_domain}. We visually present this method in Fig.~\ref{fig:noesis-with-common-lora}.

Concretely, this corresponds to a modification of Equation~\ref{eqn:mix-lora}:
\begin{equation}
    W = W + \alpha \left(\underline{B^{(c)}}\underline{A^{(c)}} +  \sum_{k=1}^K \underline{B^{(k)}}\underline{A^{(k)}} \right).
\end{equation}

Here, \( A^{(k)} \) and \( B^{(k)} \) are the trainable parameters as explained before. \( A^{(c)} \in \mathbb{R}^{r_c \times q} \) and \( B^{(c)} \in \mathbb{R}^{p \times r_c}\) constitute the LoRA decomposition, with $r_c$ corresponding to the rank of the common adapter, which is mentioned in the first column of Table~\ref{tab:common_params}. All results in this table are obtained under $(\varepsilon=1.0, \delta=10^{-6})$-DP.

For the domain-specific parameters, in \codetp, there is a bottleneck structure in $W$ of $W_i^{d_{\text{model}} \times d_\text{ff}}$ and $W_o^{d_\text{ff}\times d_\text{model}}$, where $d_\text{model}$ is the model hidden size ($d_\text{model}=768$, $d_\text{ff}=2048$ for \codetp) and we have one LoRA for each matrix.

\subsection{\tool Hyperparameter Details} \label{app:hyperparameters}
\noindent \textbf{Batch Size.} Keeping the ratio of compute to memory optimal, we run all experiments with batch size $N_b=64$ on nodes with $4\times$ 24gb-RAM GPUs. The first stage of Algorithm~\ref{alg:mixed_priv_dpsgd} is trained with batch size 96, as more memory is available when the domain experts are not yet used. One notable exception is the pt120 experiment in Table~\ref{tab:common_params}, which has only batch size 64 due to the large memory requirements of (gradients and activations for) the prompt tokens.

\begin{wraptable}{l}{6.8cm}
\vspace{-4.6mm}
\caption{Hyperparameter overview of the learning rate used in the second stage of Algorithm~\ref{alg:mixed_priv_dpsgd}.}
\begin{tabular}{c | l l l}
\toprule
\textbf{Learning Rate}  & \textbf{Python} & \textbf{Java} & \textbf{Go} \\
\midrule
 $10^{-4}$ & 67.65 &  61.02 &  64.90 \\ 
 $10^{-3}$ & 69.14 & 61.19 & 66.58 \\ 
 $5\cdot 10^{-3}$ & 51.17 & 46.34 & 28.96 \\
\bottomrule
\end{tabular}
\label{tab:hyperparam_lr}
\vspace{-2mm}
\end{wraptable}
\noindent
\textbf{Optimizer \& Learning Rates.} For all training algorithms, we use the AdamW optimizer~\citep{AdamW} with a learning rate $10^{-3}$, similar to prior work~\citep{codet5p}. A small sweep of learning rates in~\Cref{tab:hyperparam_lr} shows that either a smaller or larger learning rate results in lower accuracy. For scheduling, we use a linear step scheduler with 500 steps of warmup.

\noindent
\textbf{Rank Size and \# Trainable Tokens.} When performing parameter-efficient fine-tuning with LoRA, the rank we select depends on whether the respective component is trained with DP or not. For the domain adapters, we use domain experts of $r=512$, whereas for the case of common LoRA, we use $r=4$ (as shown also in Tab.~\ref{tab:common_params}). We select rank 4 of the common adapter using Table~\ref{tab:common_params}, and in earlier research phases we used a finer sweep. For selecting the rank 512 for the domain adapters, we find that a higher rank generally has better results, however, beyond rank 512, the model would take up too much GPU memory.

\noindent
\textbf{DP Epsilon and Gradient Clipping Norm.} Throughout our experiments in the evaluation, we report results for $\epsilon=1.0$. However, we have experimented with epsilon values of $\{0.01, 0.1, 1.0, 8.0, 16.0\}$. We report the best results that guarantees privacy without significant utility tradeoff (i.e., more than one percentage point of accuracy degradation).

For gradient clipping, we have experimented with values in the range of $\left[0.01, 10\right]$ in exponential steps and have used 1.0 as the value of choice. Using a value of $0.1$, for example, yields a slightly worse performance of $69.13, 61.16, 66.48$ on Python, Java, and Go, respectively, for a \tool\ pt32 model. Similarly, the rc4 model in Table~\ref{tab:common_params} would decrease in accuracy to $68.75, 60.32, 65.32$.

\noindent
\textbf{Training settings.} Our method is built with PyTorch (v2.4.1) and HuggingFace transformers (v4.45.0), with the functionality for DP handled by Opacus (v1.5.2). We train our models on 4$\times$4090 GPUs.
During training, we use a context length of 512 tokens and a learning rate of $10^{-3}$. For training under the DP guarantee, we apply gradient clipping with a norm of 1.0 and add Gaussian noise with a standard deviation, which is scaled according to the targeted privacy budget. We use the privacy-accountant of the Opacus library to calculate the noise-multiplier constant~\citep{opacus}. A private document within the dataset can be of any length. Therefore, an epoch is defined as sampling one 512-token block for each document in the dataset. This ensures optimal use of compute resources and maintains the privacy guarantee as each epoch uses a document exactly once.

\subsection{Details on the Membership Inference Attack}\label{app:mia-details}

In this section, we provide additional details on the implementation and results of the Membership Inference Attack. The attack and privacy protection goes with the following use case: $K$ institutions hold data for their own domain. For example, one institution holds Python data, another holds Java data, and a third holds Go data. A trusted server receives all the data to train with cross-domain insights, e.g., knowledge transfer. The trusted server returns to each institution the domain-specific model parts and the shared parameters, which are then used for inference. The privacy of each domain is maintained, and the model can simultaneously benefit from the shared knowledge. The process is illustrated in Fig.~\ref{fig:noesis_pipeline} (right). 

For the results, the definition of TPR and FPR are particularly relevant:
\begin{itemize}[noitemsep,topsep=0pt]
  \item TPR is the proportion of members of the training set that are identified as members, i.e., for which $S(\mathbf{s}) > \tau$.
  \item FPR is the proportion of non-members that are incorrectly identified as members, i.e., for which $S(\mathbf{s}) > \tau$.
\end{itemize}

\begin{figure}[h]
\centering
\includegraphics[width=.99\linewidth]{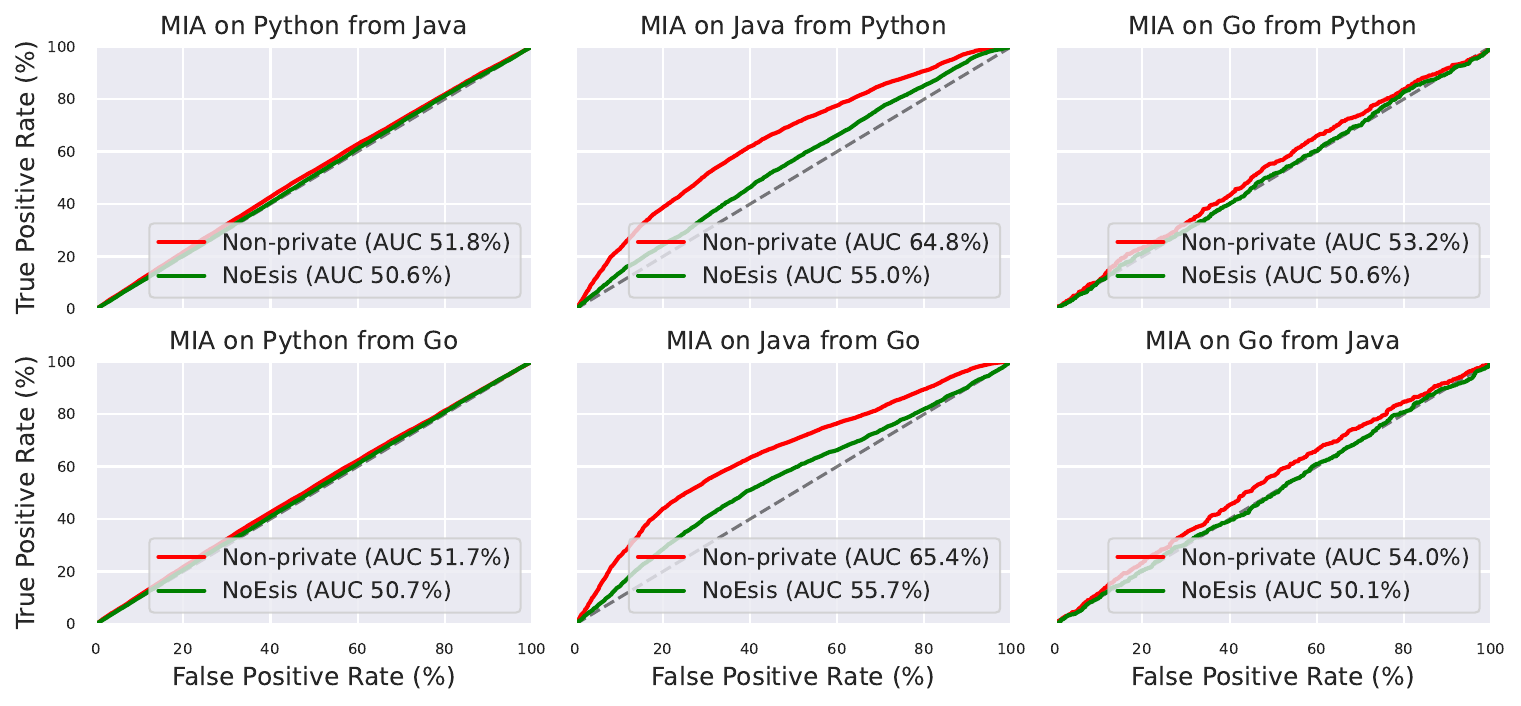}
\vspace{-0.2cm}
\caption{Visualization of results in Table~\ref{fig:roc_mia} with all combinations of source and target domains.}
\label{fig:roc_mia}
\end{figure}

From these rates, we report two aggregate metrics, both of which are common in the literature~\citep{MIA_argues_tpr}:

\begin{itemize}[noitemsep,topsep=0pt]
\item The \textbf{AUC} is a measure of the trade-off between TPR and FPR. A value of 1.0 indicates the worst possible empirical privacy protection, while 0.5 indicates good empirical privacy protection. Fig.~\ref{fig:roc_mia} shows the ROC curve for the cross-domain MIA per domain. The AUC is calculated by plotting the TPR against the FPR for all possible threshold settings and measuring the area under this curve. This corresponds to the \textsc{metrics.auc} method in~\citet{sklearn}.
\item The \textbf{TPR @1} reflects the attack accuracy rate when making a minimal number of false positives, i.e., 1\% FPR. For this evaluation, we find the threshold $\tau$ where the FPR is 1\% and report the TPR. If such a threshold does not exist, we take a weighted average of the TPR at the nearest two $\tau$ with lower and higher FPR.
\end{itemize}

The score function $S(\mathbf{s})$ is defined as the average log-likelihood assigned to each successive token of the input text by the model. This is also named \textit{teacher forcing} in the literature. In practice, this is implemented in a sequence of 512 tokens, like the training setting:
\begin{equation} \label{eqn:mia_with_ref}
 S(\mathbf{s}) = \mathcal{L}(\mathbf{s};\mathcal{M}) = \frac{1}{L-1}\sum_{l=2}^L \log \big( \mathcal{M}(s_l | s_{1:l-1}; D) \big)
\end{equation}
where $L$ is the sequence length, 512, $\mathcal{M}$ is the autoregressive model assigning log-likelihood for token $s_l$, conditioned on tokens up to $l-1$, $s_{1:l-1}$ and having been trained on dataset $D$. The threshold $\tau$ is varied across all possible scores when evaluating the ROC curve. The input tokens $s_{1:l-1}$ are from the training and test set as indicated in~\Cref{tab:datasets}. The studied Mix-LoRA model is trained with SGD, like the non-private result in Fig.~\ref{fig:bridge}.

Extending the results of Fig.~\ref{fig:roc_mia}, we provide the ROC-AUC curves for all the combinations of Python, Java, and Go source and target languages in Fig.~\ref{fig:roc_mia} below. We witness that for \tool, the TPR is closer to the random chance line, which is when the TPR equals the the FPR.

\newpage
\section{\tool Operational Details}

\begin{wrapfigure}{hr}{0.52\textwidth}
\begin{minipage}{0.52\textwidth}
\vspace{-10mm}
\begin{algorithm}[H]
\caption{Stepwise Training of \tool}
\label{alg:mixed_priv_dpsgd}
\begin{algorithmic}
\STATE {\bfseries Input:} Pre-trained model $\theta_0$, training sets $\{ D_k \}_{k=1}^K$ for K domains, collectively named $D$ with $N$ documents, batch size $N_b$, learning rate $\eta$, number of iterations $T_1$, $T_2$, privacy budget $\varepsilon, \delta$, gradient clipping bound $C$
\STATE {\bfseries Output:} Modular LLM with privacy guarantees
\vspace{2mm}

\STATE \textit{\textbf{Stage 0: Compute noise multiplier}}
\STATE $\sigma \gets \textsc{PrivAccounting}(\varepsilon, \delta, B, N, T_1)$

\STATE \textit{\textbf{Stage 1: DP training of shared parameters}}
\STATE $P_0 \gets \text{Randomly initialize prompt tokens}$
\FOR{$t = 1$ {\bfseries to} $T_1$}
  \STATE Randomly draw batch $\mathcal{B}_t$ of size $N_b$ from $D$
  \STATE $P_{t} \gets P_{t-1} - \text{DP-SGD}(\theta_0, P_{t-1}, \mathcal{B}_t, \sigma, C)$
\ENDFOR

\STATE \textit{\textbf{Stage 2: Training of domain-specific parameters}}
\STATE $E_0 \gets \text{Randomly initialize expert adapters}$
\FOR{$t = 1$ {\bfseries to} $T_2$}
  \STATE Randomly draw batch $\mathcal{B}_t$ of size $N_b$ from $D$
  \STATE $E_{t} \gets E_{t-1} - \text{SGD}(\theta_0, P_T, E_{t-1}, \mathcal{B}_t)$
\ENDFOR
\end{algorithmic}
\end{algorithm}
\vspace{-12mm}
\end{minipage}
\end{wrapfigure}

\subsection{Algorithm}

The algorithm for training \tool\ is outlined in Algorithm~\ref{alg:mixed_priv_dpsgd}. It follows a two-stage procedure of private learning for the \textit{shared parameters} (Step \blackcircle{1}) and then non-private learning for the \textit{domain-specific adapters} (Step \blackcircle{2}). 
The parameters for the expert LoRA are collectively referred to as $E$ and comprise all $B$ and $A$ matrices in Equation~\ref{eqn:mix-lora}: $E = \{ A^{(k)}, B^{(k)} \}_{k=1}^K$. The private parameters are referred to as $P$, defined in Section~\ref{sec:method}. 
The algorithm starts by computing the noise multiplier $\sigma$ using the privacy accounting mechanism~\citep{opacus}. The dataset is augmented with domain labels, which are used for deterministic routing. In each iteration, a batch of documents is randomly drawn from all domains, and the gradients are computed for the domain-specific and shared parameters. In the case of private learning, the gradients are clipped and noised according to DP-SGD~\citep{abadi_dpsgd}. In the second stage, SGD refers to Stochastic Gradient Descent. 

\subsection{Learning Curves}
To assess the convergence of our training runs, Fig.~\ref{fig:learning_curves} displays the training loss and gradient norm during the training of the models. As expected, both the loss and gradient norm decrease during the twelve training epochs. During the first stage, shown in the top row, the empirical gradient norm before clipping takes values around 1, which also happens to be our clipping norm. We have experimented with a lower clipping norm but found that this would result in lower accuracy. While the training losses for the second stage, bottom row, seem similar, training 32 prompt tokens achieves the highest accuracy on the test set, see Table~\ref{tab:common_params}. The gradient norms are calculated for the shared parameters $P_t$ (Algorithm~\ref{alg:mixed_priv_dpsgd}) in the top row and for expert parameters $E_t$ in the bottom row. The training loss is the cross-entropy log-likelihood averaged along the sequence and averaged among all three domains.

\subsection{Randomness in DP training}
\begin{figure}[h]
\centering
\begin{minipage}[t]{0.65\textwidth}
\centering \vspace{-0.01\baselineskip}
\includegraphics[width=0.99\linewidth]{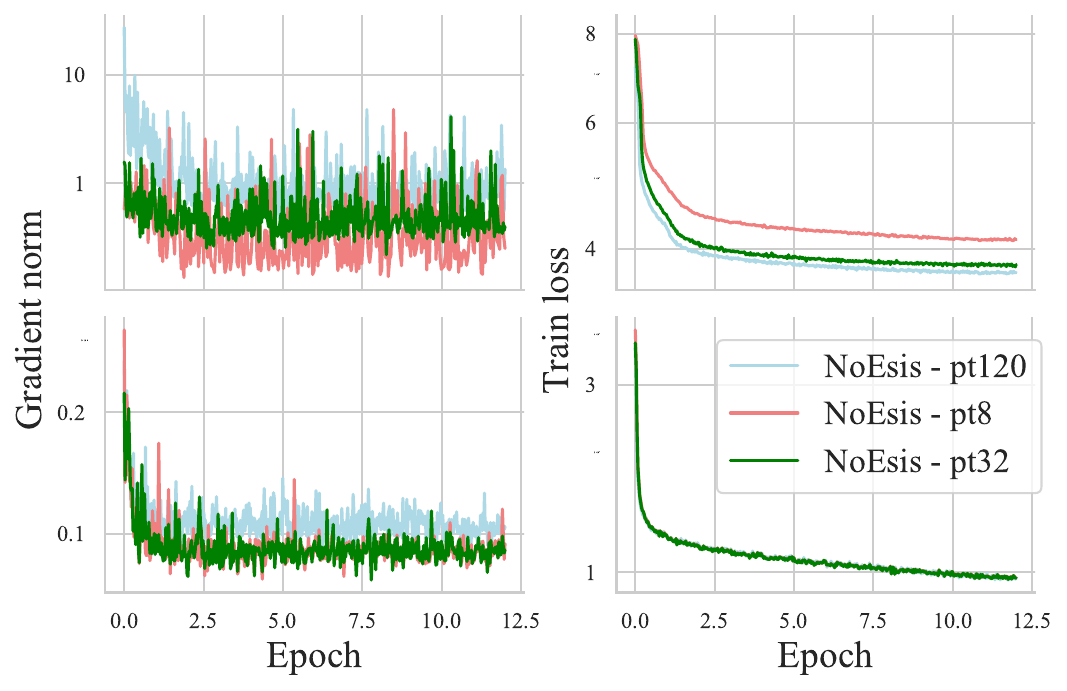}
\caption{The loss and gradient norm during training for the ablation experiment in Table~\ref{tab:ablation}. The bottom row illustrates the second training stage. While the training losses appear similar, training with 32 prompt tokens achieves the highest test accuracy.}
\label{fig:learning_curves}
\end{minipage}\hfill
\begin{minipage}[t]{0.32\textwidth}
\vspace{2mm}
\centering
\centering \vspace{-0.01\baselineskip}
\includegraphics[width=0.99\linewidth]{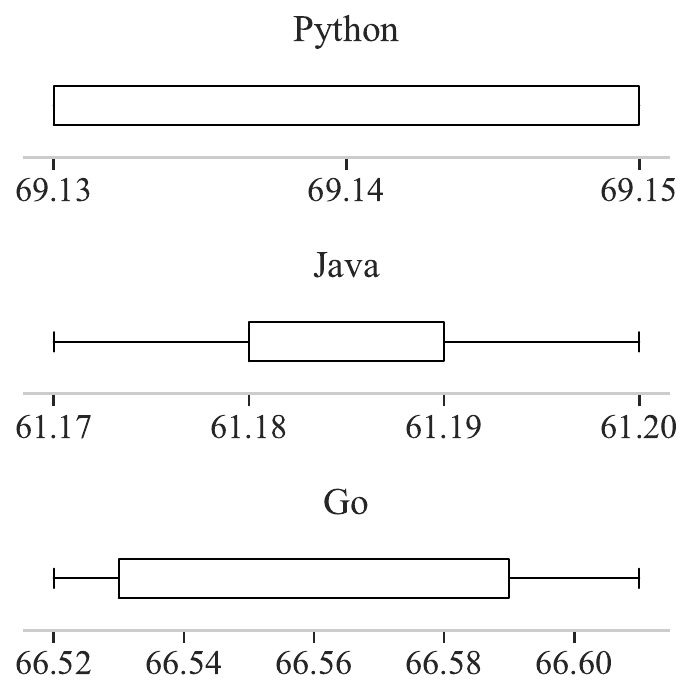}
\caption{Illustrating the stochasticity due to training \tool\ with DP-SGD. Python, having the most data, shows the least variability, while Go, as the scarce domain, exhibits greater variability in accuracy.}
\label{fig:dp_random_seeds}
\end{minipage}
\end{figure}

To explore the stochasticity introduced by training with a differentially private algorithm, we retrain \tool\ using five different random seeds. By running multiple training sessions with other seeds for the additive noise, we can better understand the range of possible outcomes. The results of these experiments are summarized by the boxplots in~\Cref{fig:dp_random_seeds}. The box edges indicate the lower and upper quartile and the whiskers correspond to the minimum and maximum values among the five random seeds. 

\subsection{Qualitative Examples of Mispredictions}\label{app:examples}
In this analysis section, we provide qualitative examples of mispredictions made by the model. These examples indicate the differences in performance between the ShareNothing model and \tool and highlight the benefit of knowledge transfer. We use color coding to demonstrate the correctness of the predictions: \textcolor{red}{\textbf{red}} indicates that both models predict the token incorrectly, \textcolor{blue}{\textbf{blue}} indicates that ShareNothing is correct but \tool is not, and \textcolor{green}{\textbf{green}} indicates that ShareNothing is incorrect but \tool gives the correct prediction. The datasets in \codexglue\ are provided stripped and tokenized, so syntax highlighting and indentation are not available.

\newpage
\begin{center}\fbox{\begin{minipage}{45em}
\textbf{  Example from Python: } \\
\tt

\noindent from  \_\_ future \_\_  import  unicode \_ l iterals ,  division ,  absolute \_ import  

  import  time  

  import  logging  

  from  collections  import  deque  

  try  :  

  from \textcolor{red}{User} Dict  import \textcolor{red}{Dict} \textcolor{red}{Mixin}  

  except  ImportError  :  

  from \textcolor{red}{collections}  import  Mapping  as  Dict Mixin  

  import  six  

  from  six  import  iteritems  

  from \textcolor{green}{six} .  moves  import  c Pick le  

  class  Base Counter  (  object  )  :  

  def  \_\_ init \_\_  (  self  )  :  

  raise  NotImplementedError  

  def  event \textcolor{green}{(}  self ,  value  =  1  )  :  

  raise  NotImplementedError  

  def \textcolor{red}{value} \textcolor{red}{(}  self , \textcolor{green}{value}  )  :  

  raise  NotImplementedError  

  @  property  

  def \textcolor{red}{avg}  (  self  )  :  

  raise  NotImplementedError  

 \textcolor{red}{@}  property  

  def \textcolor{red}{sum}  (  self  )  :  

  raise  NotImplementedError  

 \textcolor{green}{def} \textcolor{red}{empty}  (  self  )  :  

  raise  NotImplementedError  

  class  Total \textcolor{red}{Counter}  (  Base Counter  )  :  

  def  \_\_ init \_\_  (  self  )  :  

 \textcolor{red}{self} .  cnt  =  0  

  def  event  (  self ,  value  =  1  )  :  

  self .  cnt  +=  value  

  def  value  (  self ,  value  )  :  

  self .  cnt  = \textcolor{red}{value}  

  @  property  

  def  avg  (  self  )  :  

  return  self .  cnt  

  @  property  

  def  sum  (  self  )  :  

  return  self .  cnt

  {def}  empty  (  self  )  :  

  return  self .  cnt \textcolor{red}{==}  0  

  class  Average Window Counter  (  Base Counter  )  :  

  def  \_\_ init \_\_  (  self ,  window \_ size  =  300  )  :  

  self .  window \_ size  =  window \_ size  

  self .  values  =  deque  (  maxlen  =  window \_ size  )  

  def  event  (  self ,  value  =  1  )  :  

  self .  values .  append  (  value  )  

  value  = \textcolor{red}{event}  

  @  property  

  def  avg  (  self  )  :  

  return \textcolor{green}{self} .  sum  /  len  (  self .  values  )

\end{minipage}}\end{center}

\newpage
\begin{center}\fbox{\begin{minipage}{45em}
\textbf{  Example from Java: } \\
\tt

\noindent package  org .  odd \textcolor{red}{job} . \textcolor{red}{values}  ;  import  java .  util .  LinkedHashMap  ;  import  java .  util .  Map  ;  import  org . \textcolor{green}{apache} .  commons .  bean utils .  expression .  Default Resolver  ;  import  org .  apache .  commons .  bean utils . \textcolor{red}{expression} .  Resolver \textcolor{green}{;}  import  org .  odd job .  a \textcolor{red}{ro} oa . \textcolor{green}{A} ro \textcolor{red}{oa} \textcolor{red}{Exception}  ;  import  org .  odd job .  a ro oa . \textcolor{green}{A} ro \textcolor{red}{oa} \textcolor{red}{Value}  ;  import  org .  odd job .  a ro oa . \textcolor{red}{reflect} .  A ro oa \textcolor{red}{Property} Exception  ;  import  org .  odd job .  a ro oa .  reflect .  PropertyAccessor  ;  import  org .  odd job .  framework . \textcolor{red}{Simple} Job  ;  public  class \textcolor{red}{Set} Job \textcolor{red}{extends}  Simple Job  \{  private  final  Map  <  String ,  A ro \textcolor{red}{oa} \textcolor{red}{Value}  >  values  =  new  LinkedHashMap  <  String ,  A ro oa Value  >  (  )  ; \textcolor{red}{public}  void  setValues  (  String \textcolor{red}{name} , \textcolor{red}{A} ro \textcolor{green}{oa} Value  value  )  \{  values .  put  (  name ,  value  )  ;  \}  protected  int \textcolor{red}{execute}  (  )  throws  Exception  \{ \textcolor{red}{for}  (  Map .  Entry  <  String ,  A ro oa Value  >  entry  :  values .  entrySet  (  )  )  \{  String \textcolor{green}{name}  =  entry .  getKey  (  )  ;  A ro \textcolor{green}{oa} Value  value  =  entry .  getValue  (  )  ;  logger  (  ) .  info  (  " Setting  ["  +  name  + \textcolor{red}{"]}  =  ["  +  value  +  "]"  )  ; \textcolor{red}{setProperty}  (  name ,  value  )  ;  \}  return  0  ;  \}  private  void  setProperty  (  String  property ,  A ro \textcolor{green}{oa} Value  value  ) \textcolor{red}{throws}  A ro \textcolor{red}{oa} Property Exception  \{  Resolver  resolver  = \textcolor{red}{new}  Default Resolver  (  )  ;  String  comp Name  =  resolver . \textcolor{red}{next} \textcolor{red}{(}  property  )  ;  String \textcolor{red}{property} Expression  =  resolver .  remove \textcolor{red}{(} \textcolor{green}{property}  )  ;  if  (  property Expression  ==  null  )  \{ \textcolor{green}{throw}  new  A ro oa Exception  (  ""  )  ;  \}  Object  component  = \textcolor{red}{getA} ro \textcolor{red}{oa} \textcolor{red}{Session}  ( \textcolor{green}{)} . \textcolor{red}{getBean} Registry  ( \textcolor{green}{)} .  lookup  ( \textcolor{red}{comp} Name  )  ;  if  (  component  ==  null  )  \{  throw  new  A ro oa \textcolor{red}{Exception}  (  ""  + \textcolor{red}{comp} Name \textcolor{green}{+}  "]"  )  ;  \}  PropertyAccessor  property Accessor  =  getA ro \textcolor{green}{oa} \textcolor{green}{Session}  (  ) .  get Tools  ( \textcolor{red}{)} .  getProperty Accessor  (  )  ;  property Accessor  =  property Accessor .  accessor With Conversions  ( \textcolor{red}{getA} ro \textcolor{green}{oa} Session  (  ) . \textcolor{red}{get} \textcolor{red}{Tools}  (  ) .  getA \textcolor{green}{ro} \textcolor{red}{oa} \textcolor{red}{Converter}  (  )  )  ;  property Accessor .  setProperty  (  component , \textcolor{red}{property} Expression \textcolor{red}{,}  value  )  ;  \}  \}      \textcolor{red}{} \textcolor{red}{} \textcolor{red}{} 
 \\ \\ 
\textcolor{red}{}  package  org .  rub \textcolor{red}{ype} ople . \textcolor{red}{r} dt . \textcolor{red}{internal} . \textcolor{red}{ui} \textcolor{red}{.}  util  ;  import  java .  util .  Comparator  ;  import  java .  util .  HashSet  ;  import  java .  util .  Set  ;  import  java .  util .  Vector  ;  import  org . \textcolor{red}{ec} lipse . \textcolor{red}{j} face .  util .  Assert  ;  import  org .  ec lipse .  j face . \textcolor{red}{view} ers .  I Label Provider  ;  import  org .  ec lipse .  sw t . \textcolor{red}{S} WT  ;  import  org .  ec lipse .  sw t .  events .  Dis pose Event  ;  import  org .  ec lipse .  sw t .  events .  Dis pose Listener  ;  import  org .  ec lipse .  sw t . \textcolor{green}{events} .  Selection Listener  ;  import  org .  ec lipse .  sw t . \textcolor{red}{graphics} . \textcolor{red}{Image}  ;  import  org .  ec lipse .  sw t . \textcolor{red}{layout} .  Grid Data  ;  import  org .  ec lipse .  sw t .  layout .  Grid Layout  ;  import  org .  ec lipse .  sw t .  widgets . \textcolor{green}{Composite}  ;  import  org .  ec lipse .  sw t .  widgets . \textcolor{red}{Event}  ;  import  org .  ec lipse .  sw t .  widgets .  Table  ;  import  org .  ec lipse .  sw t .  widgets .  Table Item  ;  public  class  Filter edList \textcolor{red}{extends}  Composite  \{  public \textcolor{red}{interface} \textcolor{red}{Filter} \textcolor{red}{Matcher}  \{  void  setFilter  ( \textcolor{red}{String}  pattern ,  boolean \textcolor{red}{ignoreCase} ,  boolean \textcolor{red}{ignore} W ild Cards  )  ;  boolean  match  (  Object \textcolor{red}{element}  )  ; \textcolor{red}{\}}  private  class  Default \textcolor{green}{Filter} \textcolor{green}{Matcher}  implements  Filter Matcher  \{ \textcolor{red}{private}  String Matcher  f \textcolor{red}{Matcher}  ; \textcolor{green}{public}  void  setFilter \textcolor{red}{(}  String  pattern \textcolor{red}{,}  boolean  ignoreCase ,  boolean  ignore W ild Cards  )  \{  f Matcher  =  new  String Matcher  (  pattern  +  '*' ,  ignoreCase ,  ignore W ild Cards  )  ;  \}  public  boolean \textcolor{red}{match}  ( \textcolor{green}{Object}  element  )  \{  return  f Matcher .  match  (  f Renderer . \textcolor{red}{getText}  ( \textcolor{red}{element}  ) \textcolor{red}{)}  ;  \}  \} 

\end{minipage}}\end{center}

\newpage
\begin{center}\fbox{\begin{minipage}{45em}
\textbf{  Example from Go: } \\
\tt

\noindent func  Average Color  ( \textcolor{red}{colors} ... \textcolor{green}{Color}  )  (  color  Color \textcolor{red}{)}  \{  var \textcolor{red}{(} \textcolor{red}{x} ,  y \textcolor{red}{float} 64  \\
 \textcolor{red}{hue} \textcolor{green}{,} \textcolor{red}{sat} , \textcolor{red}{b} \textcolor{red}{ri} ,  kel \textcolor{red}{int}  \\
  )  \\
 \\
  //  Sum  s ind / \textcolor{red}{cos} d  for  h ues \textcolor{red}{for}  \_ ,  c  :=  range  colors  \{  //  Convert \textcolor{red}{hue}  to  degrees \textcolor{red}{h}  := \textcolor{red}{float} 64  ( \textcolor{red}{c} \textcolor{green}{.} \textcolor{red}{H} ue  ) \textcolor{red}{/} \textcolor{red}{float} 64  ( \textcolor{red}{math} . \textcolor{red}{Max} Uint 16  )  *  360 \textcolor{red}{.} 0  \\
 \\
  x \textcolor{red}{+=} \textcolor{red}{math} . \textcolor{red}{C} os  ( \textcolor{red}{h}  /  180 . 0  *  math .  Pi  )  \\
  y  +=  math .  S \textcolor{red}{in}  (  h  /  180 . 0  *  math .  Pi  )  \\
  sat \textcolor{red}{+=}  int  (  c .  Sat uration  )  \\
  b \textcolor{red}{ri}  +=  int  (  c .  B \textcolor{green}{right} ness  )  \\
  kel \textcolor{red}{+=}  int  (  c . \textcolor{red}{K} el \textcolor{red}{vin}  )  \\
  \}  \\
 \\
 \textcolor{green}{//}  Average  s ind / \textcolor{red}{cos} d  x  /=  float 64  ( \textcolor{red}{len}  (  colors  )  )  \\
 \textcolor{red}{y}  /=  float 64  (  len  (  colors  )  )  \\
 \\
  // \textcolor{red}{Take} \textcolor{red}{atan} 2 \textcolor{red}{of} \textcolor{red}{aver} aged \textcolor{red}{hue} \textcolor{red}{and}  convert \textcolor{green}{to} \textcolor{red}{uint} 16 \textcolor{red}{scale} \textcolor{red}{hue} \textcolor{red}{=} \textcolor{red}{int}  (  (  math . \textcolor{red}{At} an \textcolor{red}{2}  ( \textcolor{red}{y} ,  x  )  * \textcolor{red}{180} \textcolor{red}{.} 0  / \textcolor{red}{math} .  Pi  )  / \textcolor{red}{360} . 0  * \textcolor{red}{float} 64  (  math . \textcolor{green}{Max} Uint 16  )  )  \\
  sat  /=  len  (  colors  )  \\
  b \textcolor{red}{ri}  /= \textcolor{red}{len}  (  colors  )  \\
  kel  /= \textcolor{red}{len}  (  colors  )  \\
 \\
  color . \textcolor{red}{H} ue  =  uint 16  (  hue  )  \\
  color .  Sat uration  =  uint 16  (  sat  )  \\
  color .  B right ness  =  uint 16  (  b ri  )  \\
  color .  K el vin  =  uint 16  (  kel  )  \\
 \textcolor{red}{\\}  return \textcolor{green}{color}  \\
  \}  \\
  func  Color \textcolor{red}{Equal}  (  a ,  b  Color  )  bool  \{  return  a \textcolor{red}{.} \textcolor{red}{H} ue  ==  b .  H ue  \&\&  a .  Sat uration  ==  b .  Sat uration \textcolor{green}{\&\&}  a .  B right ness  ==  b .  B right ness  \&\&  a .  K el vin  ==  b .  K el vin  \\
  \}  \\
  func  ( \textcolor{red}{s}  *  Subscription  ) \textcolor{red}{notify}  (  event  interface  \{  \} \textcolor{green}{)}  error  \{  timeout  :=  time . \textcolor{red}{After}  (  Default Timeout  )  \\
  select  \{  case  <-  s .  quit Chan  :  Log \textcolor{green}{.}  Debugf  (  "  " ,  s . \textcolor{red}{id}  )  \\
  return  Err Closed  \\
  case  s . \textcolor{red}{events}  <-  event  :  return \textcolor{red}{nil}  \\
  case  <-  timeout  : \textcolor{red}{Log} .  Debugf  (  "  " ,  s .  id  )  \\
  return  Err \textcolor{green}{Timeout}  \\
  \}  \\
  \} 
\end{minipage}}\end{center}
\newpage

\end{document}